\documentclass[aps,prd,preprint,titlepage,nofootinbib,amsmath,amsfonts]{revtex4-2}
\usepackage[english]{babel}
\usepackage[colorlinks,allcolors=blue]{hyperref}
\usepackage[letterpaper,top=2cm,bottom=2cm,left=3cm,right=3cm,marginparwidth=1.75cm]{geometry}

\usepackage[mathscr]{euscript}
\usepackage[textsize=scriptsize,textwidth=2.5cm]{todonotes}
\usepackage{amssymb,amsmath,latexsym,bm,amsfonts,mathtools}
\usepackage{graphicx}
\usepackage{ulem}
\usepackage{dsfont}
\usepackage{color}
\usepackage{indentfirst}
\usepackage{subfig}
\usepackage{comment}
\usepackage{caption} %customises captions
\usepackage{floatrow}

\usepackage{multirow}
\usepackage{booktabs}
\usetikzlibrary{positioning, shapes.geometric, patterns, graphs, decorations.pathmorphing}
\usepackage{relsize}
\usepackage{graphicx}
\usepackage{calc}

\begin{document}
\title{Cosmological Inflation and Dark Sector from 11D Supergravity }

\author{Jiaming Shi}
\email{shijiaming@ucas.ac.cn}
\affiliation{$^{1}$School of Fundamental Physics and Mathematical Sciences, Hangzhou Institute for Advanced Study,
University of Chinese Academy of Sciences, Hangzhou 310024, China}
\affiliation{$^{2}$School of Physical Sciences, University of Chinese Academy of Sciences, No.19A Yuquan Road, Beijing 100049, China}

\author{Zehua Xiao\footnote{Co-first author}
\footnote{Corresponding author}}
\email{xiaozh20@mails.tsinghua.edu.cn}
\affiliation{Yau Mathematical Sciences Center, Tsinghua University, Beijing 100084, China}

\begin{abstract}
We explore compactifications of the form $\mathcal{M}^{3,1}\times T_{q_{1}}^{2}\times T_{q_{2}}^{2}\times T_{q_{3}}^{2}\times S^{1}$ in the framework of 11D supergravity. By imposing suitable gauge conditions and boundary conditions, we find that the FRW universe with four extended spacetime dimensions and seven extremely small compactified spatial dimensions emerges as a solution for the 11D supergravity. These specific compactification methods can produce cosmological inflation that aligns with the observational constraints set by the 2021 BICEP/Keck \cite{BICEP:2021xfz} and Planck 2018 results \cite{Planck:2018jri}.  In the cosmological inflation models we construct, the inflaton can be interpreted as the conformal vibrations of extra dimensions with a size around $10^{5}$ times the reduced Planck length. Additionally, we offer the theoretical predictions for the mass of the inflaton, and the tree-level Newton's gravity law between two massive point particles surrounded by a spherically symmetric distribution of the inflaton, which can reproduce the Tully-Fisher relation and explain the flat rotation curves of galaxies.
\end{abstract}

\maketitle

\newpage

\section{Introduction}
The eleven-dimensional supergravity is widely believed to be the tree-level low-energy effective theory of M-theory. As an ultraviolet complete theory, M-theory possesses inherent advantages in addressing extremely high-energy scenarios, such as the inflationary period of the early universe. However, our understanding of M-theory remains limited, see \cite{Banks:1996vh, Polchinski:1999br, Taylor:2001vb} for remarkable progress in M-theory. A comprehensive theoretical description is still necessary for understanding physics in the extremely early universe by M-theory. From the perspective of string theory, it is conjectured that by compactifying M-theory on a special manifold of $G_{2}$ holonomy, it could yield the supersymmetric version of Standard Model of Particle Physics and General Relativity \cite{Acharya:2007rc, Kane:2011kj}. This conjecture has generated some discussions on M-theory cosmology in recent years, particularly regarding $G_{2}$ compactification, see \cite{Kane:2019nod, Brustein:2002xf, Coley:2002xb} for examples. While this paper does not aim to fully realize the ambitious goals of M-theory cosmology, it seeks to explore other potential ways for dimensional reduction that could lead to cosmological inflation in the early universe.

In contrast to M-theory, its low-energy effective theory, eleven-dimensional supergravity, has been more extensively studied. Comprehensive reviews, such as \cite{Freedman:2012zz, Green:2012pqa}, provide a valuable resource for understanding this field. It is believed that eleven-dimensional supergravity serves as the mother theory for lower-dimensional supergravity theories. Numerous papers have delved into various lower-dimensional supergravity theories and their cosmological models \cite{DEath:1996ejf, Yamaguchi:2011kg, Ferrara:2016ajl, Lauria:2020rhc, Ferrara:2013rsa, Halyo:1996pp, Kallosh:2010xz}, including discussions on no-scale inflation as discussed in references \cite{Ellis:2013xoa, Ellis:2013nxa, Wu:2024lrw, Wu:2022kew}. However, cosmological inflation models based on the original eleven-dimensional supergravity remain rare. Early discussions have touched upon the inflationary scenario involving the compactification of seven dimensions as $S^{7}$ \cite{Hawking:1998ub, Bremer:1998zp}. Other approaches from M-theory to string cosmology such as Calabi-Yau compactification $CY^{3}\times S^{1}$ were studied in \cite{Billyard:1999dg, Cicoli:2008gp}.

After inflation, dark matter plays a vital role in cosmic expansion and galaxy formation which can be interpreted as a scalar field, specifically the axion or dilaton field \cite{Svrcek:2006yi,Schive:2014dra,Marsh:2015daa, Stadnik:2015kia, Tenkanen:2019aij,Appelquist:2024koa, OHare:2024nmr}. The dark scalar field non-minimally coupled to gravity was studied in Refs. \cite{Cosme:2018nly,Alonso-Alvarez:2018tus,Shi:2019hwe}. When considering galactic scales, a scalar field is interpreted as a condensate of ultra-light bosonic dark matter (UBDM) that interacts with general relativity, adhering to the Klein-Gordon equations. In the non-relativistic realm, the equations of motion reduce to the Schr\"{o}dinger-Poisson system. It is postulated that, at galactic scales, this dark matter forms power-law matter density profiles described by $\rho=\rho_{0}r^{-\alpha}$ \cite{Pierobon:2023ozb, Eggemeier:2024fzs, Bertschinger:1985pd} or the Navarro-Frenk-White matter density profiles \cite{Navarro:1995iw, Primack:1998sa}
\begin{align}
\rho(r)=\frac{\rho_{0}}{r/r_{s}(1+r/r_{s})^{2}},\label{density0}
\end{align}
where $r_{s}$ is the characteristic scale radius. On the other hand, MOdified Newtonian Dynamics (MOND) \cite{Milgrom:1983zz,Milgrom:1983ca,Bekenstein:1984tv,Milgrom:1986ib} is a possible and promising alternative to the dark matter problem which can explain the flat rotation curves of galaxies and the Tully-Fisher relation $v^{4}\propto M$ \cite{Tully:1977fu,McGaugh:2000sr,Torres-Flores:2011bna,Alestas:2021nmi}, where $v$ is the rotation speed of a galaxy and $M$ is the total baryonic mass (stars plus gas) in the galaxy. However, it is challenged by the observation of Bullet cluster \cite{Clowe:2006eq} since the modification to dynamics can not reveal the dark matter distribution. To consider the advantages of the MOND theory, we expect that some hidden matter could induce results that are similar to the implications of MOND.

Another mystery in the dark sector of the universe is dark energy \cite{SupernovaSearchTeam:1998fmf,SupernovaCosmologyProject:1998vns}. The dark energy with negative pressure dominates the evolution of our current universe and explains the accelerated expansion of the universe. In the standard cosmological model ($\Lambda$CDM model), the cosmological constant $\Lambda$ as the dark energy contributes about 70\% of the energy density of the current universe. Unfortunately, the physical interpretation of the dark energy suffers from the cosmological constant problem or
the fine-tuning problem since the value of the energy density of cosmological constant $\rho_\Lambda\simeq 10^{-120}m_p^4$, where  $m_p \sim 10^{18} \mathrm{GeV}$ is a Planck mass scale  \cite{Carroll:1991mt,Martin:2012bt}.  This problem may be alleviated by the mechanism of small-scale spacetime fluctuations \cite{Wang:2019mee} or spacetime foam \cite{Carlip:2018zsk}. In fact, dark energy is not necessary to be a cosmological constant. Recent observations from the Dark Energy Spectroscopic Instrument (DESI), along with other datasets including CMB, PantheonPlus, Union3, and DESY5, support the time-varying dark energy equation of state \cite{DESI:2024mwx}.

In this study, we use the tree-level eleven-dimensional supergravity, which we will abbreviate as ``11D supergravity", as the starting point. The action of the 11D supergravity is given by \cite{Freedman:2012zz}
\begin{equation}
\begin{split}
S&=\frac{m_{p(11)}^{9}}{2}\int d^{11}x\sqrt{-G_{11}}\Bigg[e^{aM}e^{bN}R_{MNab}(\omega)-\bar{\psi}_{A}\Gamma^{ABC}D_{B}\left(\frac{\omega+\hat{\omega}}{2}\right)\psi_{C}
-\frac{1}{24}F^{ABCD}F_{ABCD}\\
&-\frac{\sqrt{2}}{192}\bar{\psi}_{M}
\left(\Gamma^{ABCDMN}+12\Gamma^{AB}g^{CM}g^{DN}\right)\psi_{N}\left(2F_{ABCD}+\frac{3}{2}\sqrt{2}\bar{\psi}_{[A}\Gamma_{BC}\psi_{D]}\right)\\
&-\frac{2\sqrt{2}}{(144)^{2}}\varepsilon^{A'B'C'D'ABCDMNP}F_{A'B'C'D'}F_{ABCD}A_{MNP}\Bigg],\label{sugra}
\end{split}
\end{equation}
where the indices $A,B,C,\dots=0,1,2,\cdots,10$ and $m_{p(11)}$ is the 11 dimensional reduced Planck mass.  $e_{M}^{a}$ is the frame fields with $a=0,1,2,\cdots,10$. There are 11D graviton $G_{AB}$, 11D gravitino $\psi_{A}$ and 3-form $A_{MNP}$ as elementary fields in 11D supergravity. The field strength $F_{ABCD}$ for the 3-form is defined as $F=dA$. The 3-form possesses a gauge invariance given by $A'=A+d\theta$, in which $\theta_{MN}$ serves as a 2-form gauge parameter. The 32 dimensional gamma matrices are $\Gamma^{A_{1}A_{2}\cdots A_{k}}=\Gamma^{[A_{1}}\Gamma^{A_{2}}\cdots\Gamma^{A_{k}]}$. $\varepsilon^{A'B'C'D'ABCDMNP}$ is the totally antisymmetric rank 11 tensor with $\varepsilon^{012\cdots10}=-1$. The covariant derivative is defined as
\begin{align}
D_{M}\left(\tilde{\omega}\right)&=\nabla_{M}+\frac{1}{4}\tilde{\omega}_{Mab}\Gamma^{ab},
\end{align}
where the connection $\tilde{\omega}$ can be
\begin{align}
\omega_{Mab}&=\omega_{Mab}(e)+K_{Mab},\\\hat{\omega}_{Mab}&=\omega_{Mab}(e)-\frac{1}{4}\left(\bar{\psi}_{M}\Gamma_{b}\psi_{a}-\bar{\psi}_{a}\Gamma_{M}\psi_{b}+\bar{\psi}_{b}\Gamma_{a}\psi_{M}\right),\\\omega_{M}^{\ \ ab}(e)&=2e^{N[a}\nabla_{[M}e_{N]}^{\ \ \ b]}-e^{N[a}e^{b]P}e_{Mc}\nabla_{N}e_{P}^{\ \ c},\\K_{Mab}&=-\frac{1}{4}\left(\bar{\psi}_{M}\Gamma_{b}\psi_{a}-\bar{\psi}_{a}\Gamma_{M}\psi_{b}+\bar{\psi}_{b}\Gamma_{a}\psi_{M}\right)+\frac{1}{8}\bar{\psi}_{N}\Gamma_{\ \ \ \ Mab}^{NP}\psi_{P}.
\end{align}
 The action \eqref{sugra} is invariant under supersymmetry transformation
\begin{align}
\delta e_{M}^{a}&=\frac{1}{2}\bar{\epsilon}\Gamma^{a}\psi_{M},\label{susytran1}\\
\delta\psi_{M}&=D_{M}\left(\hat{\omega}\right)\epsilon+\frac{\sqrt{2}}{288}\left(\Gamma_{\ \ \ \ \ \ \ M}^{ABCD}-8\Gamma^{BCD}\delta_{M}^{A}\right)\left(F_{ABCD}+\frac{3\sqrt{2}}{2}\bar{\psi}_{[A}\Gamma_{BC}\psi_{D]}\right)\epsilon,\label{susytran2}\\
\delta A_{MNP}&=-\frac{3\sqrt{2}}{4}\bar{\epsilon}\Gamma_{[MN}\psi_{P]},\label{susytran3}
\end{align}
where $\Gamma^{a}, a=0,1,2,\cdots,10$ is the gamma matrix in flat eleven-dimensional tangent space. $\epsilon$, with its Dirac conjugation $\bar{\epsilon}$, is the spinor gauge parameter of eleven-dimensional supersymmetry. In this paper, we define the four-dimensional reduced Planck mass $m_{p}=\frac{1}{\sqrt{8\pi G}}$, where $G$ is the four-dimensional Newtonian gravitational constant.

In Section \ref{sectionII}, we propose that by compactifying seven dimensions in the form of $\mathcal{M}^{3,1}\times T_{q_{1}}^{2}\times T_{q_{2}}^{2}\times T_{q_{3}}^{2}\times S^{1}$, the action \eqref{sugra} can give rise to the complete evolution of the four-dimensional visible universe, encompassing matter described by the Standard Model of Particle Physics as well as dark matter and dark energy. The four-dimensional extended spacetime, denoted by $\mathcal{M}^{3,1}$, is described using the Friedmann-Robertson-Walker (FRW) metric. The gauge conditions for the gauge parameters $\theta_{MN}$ and $\epsilon$ of the 11D supergravity can be chosen to prevent the extra dimensions from decompactifying as the universe evolves. Through numerical analysis, in Section \ref{sectionIII}, we will see that a contraction of the 3 tori and an expansion of the eleventh dimension gives rise to cosmological inflation in the extended FRW universe. The inflaton is interpreted as the conformal vibrations of the extra dimensions.  Furthermore, in Section \ref{quantumEFF}, we discussed the quantum effects coming from the compactified extra dimensions, where the inflaton is supposed to be dark matter or dark energy. We will see that the leading order contribution from the four-point matter scattering amplitude results in a Yukawa-like modification to Newton's law of gravity. Furthermore, this leading order relativistic contribution can explain the Tully-Fisher relation $v^{4}\sim M$ observed in galaxies.

\section{Compactification Towards Four Dimensions}\label{sectionII}
The universe we see comprises four extended dimensions: three spatial dimensions and one temporal dimension. In the theory of eleven-dimensional supergravity with all loop-order corrections, the remaining seven dimensions of the theory should be compactified to an extremely small scale. These compactified dimensions are presumably microscopic and their vibrations are influenced by quantum effects. At the beginning of the universe, scattering processes involving gravitons, gravitinos, and 3-forms are prevalent, resulting in a boiling 11D spacetime that can not be perceived as a background of the tree-level eleven-dimensional supergravity \eqref{sugra}. In this scenario, the supergravity necessitates the inclusion of an infinite number of loop-order quantum corrections, which lie beyond our current analytical capabilities. Hence, we postulate that the inflationary period commences when the loop-order graviton scattering processes become a negligible contribution to the universe's total S-matrix, rendering all the loop-order quantum corrections of the graviton field insignificant. In our inflation model, matter fields of four-dimensional effective action are assumed temporarily inactive during the inflation period, resulting in the graviton and inflaton dominating the universe's dynamics. This assumption of the inflationary period can be interpreted as selecting suitable gauge conditions for the gauge parameters $\theta_{MN}$ and $\epsilon$ and boundary conditions for the 11D supergravity.

\subsection{A Special Case}\label{subectionA}
\indent In this subsection, we will focus on a specific dimensional reduction method from 11D to 4D. It involves seven compactified extra dimensions: 2 dimensions form a 2D torus $T^{2}_{q_{1}}$
with genus $q_{1}$, 2 dimensions form a 2D torus $T^{2}_{q_{2}}$ with genus $q_{2}$,
the other 2 dimensions form another 2D torus $T^{2}_{q_{3}}$ with genus $q_{3}$ ,
the last 1 dimension forms a circle. The 11D bulk metric is given by
\begin{align}
ds_{11}^{2} & =G_{AB}dx^{A}dx^{B}.
\end{align}
Four extended dimensions within the 11D bulk exhibit the FRW metric:
\begin{equation}
ds_{4}^{2}=g_{\mu\nu}dx^{\mu}dx^{\nu}=-dt^{2}+a^{2}(t)\delta_{ij}dx^{i}dx^{j},
\end{equation}
where $\mu,\nu=0,1,2,3$. The metrics of the three 2D tori are given as follows
\begin{equation}
\begin{split}
ds_{2(1)}^{2} & =\Phi(x^{\mu})g_{a_{1}b_{1}}(x^{a_{1}})dx^{a_{1}}dx^{b_{1}},\\
ds_{2(2)}^{2} & =h_{a_{2}b_{2}}(x^{a_{2}})dx^{a_{2}}dx^{b_{2}},\\
ds_{2(3)}^{2} & =\Phi^{2}(x^{\mu})\gamma_{a_{3}b_{3}}(x^{a_{3}})dx^{a_{3}}dx^{b_{3}},\label{2dmetric}
\end{split}
\end{equation}
where $a_{1},b_{1}=4,5$, $a_{2},b_{2}=6,7$, and $a_{3},b_{3}=8,9$. The scalar
$\Phi(x^{\mu})>0$ is the scale function of the tori. It is dynamical, allowing the tori to conformally transform into larger or smaller sizes by changing the value of the function $\Phi(x^{\mu})$. The metrics above with scale function $\Phi(x^{\mu})=1$ can be used to describe the standard sizes of these three tori.

The 11th dimension is compactified into a circle by
\begin{equation}
ds_{1}^{2}=\Phi^{-6}(x^{\mu})dx^{10}dx^{10}.\label{11thmetric}
\end{equation}

Hence, the metric of the 11D supergravity can be written as
\begin{equation}
ds_{11}^{2}=ds_{4}^{2}+ds_{2(1)}^{2}+ds_{2(2)}^{2}+ds_{2(3)}^{2}+ds_{1}^{2}.\label{11dmetric}
\end{equation}

The 11D Ricci scalar of $G_{AB}$ after dimensional reduction is given by
\begin{equation}
R_{(11)}=R_{(4)}-\frac{23}{2\Phi^{2}}\nabla_{\mu}\Phi\nabla^{\mu}\Phi+\frac{R_{T_{q_{1}}^{2}}}{\Phi}+R_{T_{q_{2}}^{2}}+\frac{R_{T_{q_{3}}^{2}}}{\Phi^{2}},\label{11dRicci}
\end{equation}
where $R_{(4)}$ is the Ricci scalar of the metric $g_{\mu\nu}$ and $R_{T_{q_{i}}^{2}}, i=1,2,3$
are the Ricci scalars of the metric $g_{a_{1}b_{1}},h_{a_{2}b_{2}},\gamma_{a_{3}b_{3}}$
respectively. The non-zero equations of motion of $G_{AB}$ are provided by Einstein's equations
\begin{align}
\mathcal{G}_{\mu\nu}-\frac{23}{2\Phi^{2}}\nabla_{\mu}\Phi\nabla_{\nu}\Phi-\frac{1}{2}g_{\mu\nu}\left(\frac{R_{T_{q_{1}}^{2}}}{\Phi}+R_{T_{q_{2}}^{2}}+\frac{R_{T_{q_{3}}^{2}}}{\Phi^{2}}-\frac{23}{2\Phi^{2}}\nabla_{\rho}\Phi\nabla^{\rho}\Phi\right)&=\frac{1}{m_{p(11)}^{9}}T_{\mu\nu}^{(A,\psi)},\label{4dEinstein}\\
\mathcal{G}_{a_{1}b_{1}}-\frac{1}{2}g_{a_{1}b_{1}}\left(\Phi R_{(4)}+\Phi R_{T_{q_{2}}^{2}}+\frac{R_{T_{q_{3}}^{2}}}{\Phi}+\nabla_{\mu}\nabla^{\mu}\Phi-\frac{25}{2\Phi}\nabla_{\mu}\Phi\nabla^{\mu}\Phi\right)&=\frac{1}{m_{p(11)}^{9}}T_{a_{1}b_{1}}^{(A,\psi)},\\
\mathcal{G}_{a_{2}b_{2}}-\frac{1}{2}h_{a_{2}b_{2}}\left(R_{(4)}+\frac{R_{T_{q_{1}}^{2}}}{\Phi}+\frac{R_{T_{q_{3}}^{2}}}{\Phi^{2}}-\frac{23}{2\Phi^{2}}\nabla_{\mu}\Phi\nabla^{\mu}\Phi\right)&=\frac{1}{m_{p(11)}^{9}}T_{a_{2}b_{2}}^{(A,\psi)},\\
\mathcal{G}_{a_{3}b_{3}}-\frac{1}{2}\gamma_{a_{3}b_{3}}\left(\Phi^{2}R_{(4)}+\Phi R_{T_{q_{1}}^{2}}+\Phi^{2}R_{T_{q_{2}}^{2}}+\Phi\nabla_{\mu}\nabla^{\mu}\Phi-\frac{27}{2}\nabla_{\mu}\Phi\nabla^{\mu}\Phi\right)&=\frac{1}{m_{p(11)}^{9}}T_{a_{3}b_{3}}^{(A,\psi)},\\
-\frac{1}{2}\left(\frac{R_{(4)}}{\Phi^{6}}+\frac{R_{T_{q_{1}}^{2}}}{\Phi^{7}}+\frac{R_{T_{q_{2}}^{2}}}{\Phi^{6}}+\frac{R_{T_{q_{3}}^{2}}}{\Phi^{8}}-\frac{6}{\Phi^{7}}\nabla_{\mu}\nabla^{\mu}\Phi-\frac{11}{2\Phi^{8}}\nabla_{\mu}\Phi\nabla^{\mu}\Phi\right)&=\frac{1}{m_{p(11)}^{9}}T_{10,10}^{(A,\psi)},
\end{align}
where $\mathcal{G}_{\mu\nu},\mathcal{G}_{a_{i}b_{i}},i=1,2,3$, are the Einstein tensor for metric $g_{\mu\nu},g_{a_{1}b_{1}},h_{a_{2}b_{2}},\gamma_{a_{3}b_{3}}$ respectively. The energy-momentum tensor, denoted as $T_{AB}^{(A,\psi)}$, is defined for the gravitino $\psi_{A}$ and 3-form $A_{MNP}$ of 11D supergravity. Note that there are gauge parameters $\theta_{MN}$ and $\epsilon$ in 11D supergravity. It is imperative to incorporate gauge conditions, see \ref{Appendix1} for more details, when solving the equations of motion for the gravitino and the 3-form. After imposing suitable gauge conditions, the solutions for the gravitino and 3-form can be derived from the equations of motion. Note that the compactification method leads to the spontaneous breaking of all the $\mathcal{N}=32$ supersymmetries of 11D supergravity.

As an example, concentrate on the trivial solutions to the equations of motion for gravitino $\psi_{A}$ and 3-form $A_{MNP}$, where the solutions are $\psi_{A}=0$ and $A_{MNP}=0$ by adopting specific gauge conditions. The physical significance of these two solutions is that both the gravitino and the 3-form remain inactive during the entire inflationary period. Consequently, the energy-stress tensor simplifies to $T_{AB}^{(A,\psi)}=0$, leaving the 11D Einstein-Hilbert action as the outcome for action \eqref{sugra}. The gauge conditions for the gravitino and 3-form give the matter coupling term of 4D effective action is $\mathcal{L}_{\text{matter}}=0$. Interestingly, one can deduce a suitable potential $V(\Phi)$ by integrating out the physics arising from various configurations of the extra dimensions within the 11D Einstein-Hilbert action
\begin{align}
    S_{EH}=\int d^{11}x\sqrt{-G_{11}}\frac{m_{p(11)}^{9}}{2}R_{(11)}.
\end{align}
Due to the block diagonal property of the 11D metric \eqref{11dmetric}, the integral over the 11D bulk can be decomposed into multiple integrals over the 4D extended spacetime and the compactified extra dimensions, corresponding to each block of the 11D metric
\begin{align}
    \int d^{11}x \sqrt{-G_{11}}[\cdots]=\int_{V_{\text{FRW}}}d^{4}x\sqrt{-g_{4}}\int_{T^{2}_{q_{1}}}d^{2}x\sqrt{g_{2}}\int_{T^{2}_{q_{2}}} d^{2}x\sqrt{h}\int_{T^{2}_{q_{3}}} d^{2}x\sqrt{\gamma}\int_{S^1} dx[\cdots].\label{integral}
\end{align}
After integrating the 11D Ricci scalar \eqref{11dRicci}, the 11 dimensional Einstein-Hilbert action is now represented by
\begin{equation}
\begin{aligned}
S_{EH}=&\int d^{4}x\sqrt{-g_{4}}\Bigg[\frac{m_{p}^{2}}{2}R_{(4)}-\frac{1}{2}\frac{23m_{p}^{2}}{2\Phi^{2}}\nabla_{\mu}\Phi\nabla^{\mu}\Phi\\
 &+\frac{4\pi(1-q_{1})m_{p}^{2}}{A_{T_{q_{1}}^{2}}\Phi}+\frac{4\pi(1-q_{2})m_{p}^{2}}{A_{T_{q_{2}}^{2}}}+\frac{4\pi(1-q_{3})m_{p}^{2}}{A_{T_{q_{3}}^{2}}\Phi^{2}}\Bigg],\label{EHaction}
\end{aligned}
\end{equation}
where we have defined the 4D reduced Planck mass $m_{p}$ by utilizing the 11D reduced Planck mass $m_{p(11)}$, the areas of the three tori of standard sizes $A_{T^{2}_{q_{i}}},i=1,2,3$, and the standard radius of the eleventh dimension $\widetilde{\mathcal{R}}_{10}$ as follows
\begin{equation}
\frac{2\pi \widetilde{\mathcal{R}}_{10}A_{T_{q_{1}}^{2}}A_{T_{q_{2}}^{2}}A_{T_{q_{3}}^{2}}m_{p(11)}^{9}}{2}=\frac{m_{p}^{2}}{2}.\label{massrelation}
\end{equation}
The last three terms in \eqref{EHaction}, which function as the cosmological inflation potential, originate from the Gauss-Bonnet theorem
\begin{align}
  \int_{T^{2}_{q_{i}}}R_{T^{2}_{q_{i}}}=8\pi(1-q_{i}),\ i=1,2,3.
\end{align}
The second term in \eqref{EHaction} is a kinetic term, which becomes more apparent through the redefinition of the field $\Phi$ as
\begin{align}
\phi & =\sqrt{\frac{23}{2}}m_{p}\ln\Phi,\\
\Phi & =\exp\left({\sqrt{\frac{2}{23}}\frac{\phi}{m_{p}}}\right).
\end{align}
Restrict $q_{1}=0,q_{2}=2,q_{3}=2$ and $2A_{T_{q_{1}}^{2}}=A_{T_{q_{2}}^{2}}=A_{T_{q_{3}}^{2}}=A_{0}$,
we derive a single-field cosmological inflation action
\begin{align}
S_{\text{eff}}=S_{EH}=\int d^{4}x\sqrt{-g_{4}}\left[\frac{m_{p}^{2}}{2}R_{(4)}-\frac{1}{2}\nabla_{\mu}\phi\nabla^{\mu}\phi-\frac{4\pi m_{p}^{2}}{A_{0}}\left(1-e^{-\sqrt{\frac{2}{23}}\frac{\phi}{m_{p}}}\right)^{2}\right],\label{EHaction1}
\end{align}
where $\phi$ is the inflaton. The effective action \eqref{EHaction1} of the 11D bulk arises from integrating out the physics of seven extra dimensions in the low-energy limit. However, it still discribes an 11D universe with four extended spacetime dimensions and seven extremely small compactified spatial dimensions. This means the extra dimensions need to be sufficiently small and in the low energy limit,   the extra dimensions does not contribute to the effective action so that the physics from the extra dimensions are hidden. It is important to note that assigning specific values to the genus $q_{1},q_{2},q_{3}$ and the areas $A_{T_{q_{1}}^{2}},A_{T_{q_{2}}^{2}},A_{T_{q_{3}}^{2}}$ essentially imposes boundary conditions on the metric $G_{AB}$. The trick presented in \eqref{integral}\eqref{EHaction} illustrates one potential method of defining the driven potential $V(\Phi(\phi))$ as
\begin{align}
V(\Phi(\phi))=\frac{4\pi m_{p}^{2}}{A_{0}}\left(1-e^{-\sqrt{\frac{2}{23}}\frac{\phi}{m_{p}}}\right)^{2},\label{specialpotential}
\end{align}
In this approach, the relative stretched sizes of the three tori are represented as $(\Phi,1,\Phi^{2})$. While the specific construction of extra dimensions is given in \eqref{2dmetric}, it should be noted that other configurations of extra dimensions are also conceivable. Before conducting numerical analysis, we intend to explore scenarios involving various sizes combinations of the three tori.

\subsection{General Case}
\indent In this subsection, we will explore the scenario involving general relative stretched sizes of the three tori, specifically represented as $(\Phi^{2s_{1}},\Phi^{2s_{2}},\Phi^{2s_{3}})$ with parameters $s_{1},s_{2},s_{3}\geq 0$.  The corresponding metrics for these three tori are provided by
\begin{align}
ds_{2(1)}^{2} &=\Phi(x^{\mu})^{2s_{1}}g_{a_{1}b_{1}}(x^{a_{1}})dx^{a_{1}}dx^{b_{1}},\\
ds_{2(2)}^{2} &=\Phi(x^{\mu})^{2s_{2}}h_{a_{2}b_{2}}(x^{a_{2}})dx^{a_{2}}dx^{b_{2}},\\
ds_{2(3)}^{2} &=\Phi(x^{\mu})^{2s_{3}}\gamma_{a_{3}b_{3}}(x^{a_{3}})dx^{a_{3}}dx^{b_{3}}.
\end{align}

Since we anticipate the presence of Einstein's Gravity in a four-dimensional extended spacetime, the 11th dimension is compacted into a circular shape as
\begin{align}
    ds^{2}_{1}=\Phi^{-2(s_{1}+s_{2}+s_{3})}dx^{10}dx^{10}.
\end{align}
Its radius is specifically defined to prevent any coupling between $\Phi$ and the 4D Ricci scalar $R_{(4)}$. The Ricci scalar in 11 dimensions, after dimensional reduction, is expressed as follows:
\begin{align}
R_{(11)}=R_{(4)}+\frac{R_{T_{q_{1}}^{2}}}{\Phi^{2s_{1}}}+\frac{R_{T_{q_{2}}^{2}}}{\Phi^{2s_{2}}}+\frac{R_{T_{q_{3}}^{2}}}{\Phi^{2s_{3}}}-\frac{\lambda}{\Phi^{2}}g^{\mu\nu}\nabla_{\mu}\Phi\nabla_{\nu}\Phi,
\end{align}
where
\begin{align}
\lambda=6\underset{i}{\sum}s_{i}^{2}+8\underset{i > j}{\sum}s_{i}s_{j},
\end{align}
with the indices $i,j=1,2,3$. Analogous to the special case presented in Subsection \ref{subectionA}, the four-dimensional effective action for 4D extended spacetime can be derived through the imposition of appropriate gauge conditions and boundary conditions. Unsurprisingly, the effective action of the 11D bulk can be written as
\begin{align}
S_{\text{eff}} = \int d^{4}x\sqrt{-g_{4}}\left[\frac{m_{p}^{2}}{2}R_{(4)}-\frac{1}{2}\frac{\lambda m_{p}^{2}}{\Phi^{2}}\nabla_{\mu}\Phi\nabla^{\mu}\Phi-U(\Phi)+\mathcal{L}_{\text{matter}}\right].\label{EHaction2}
\end{align}
By utilizing the technique demonstrated in equations \eqref{integral} and \eqref{EHaction}, the potential can be defined as follows:
\begin{align}
U(\Phi)=-\left(\frac{4\pi(1-q_{1})m_{p}^{2}}{A_{T^{2}_{q_{1}}}\Phi^{2s_{1}}}+\frac{4\pi(1-q_{2})m_{p}^{2}}{A_{T^{2}_{q_{2}}}\Phi^{2s_{2}}}+\frac{4\pi(1-q_{3})m_{p}^{2}}{A_{T^{2}_{q_{3}}}\Phi^{2s_{3}}}\right).
\end{align}
The redefinition of the field is as stated below:
\begin{align}
\phi &=\sqrt{\lambda}m_{p}\ln\left|\Phi\right|,\\
\Phi &=\pm \exp\left({\sqrt{\lambda^{-1}}\frac{\phi}{m_{p}}}\right).\label{redefine2}
\end{align}

To compare our findings with the discussions presented in subsection \ref{subectionA}, the equations written as $2A_{T_{q_{1}}^{2}}=A_{T_{q_{2}}^{2}}=A_{T_{q_{2}}^{2}}=A_{0}$ remain applicable. As the inflaton $\phi$ rolls down to a true vacuum, we can establish conditions $U(\Phi_{min})=0,U'(\Phi_{min})=0$ on the potential. The solution is given by
\begin{align}
q_{1}=\frac{\Phi_{min}^{2s_{1}-2s_{3}}(s_{2}-s_{3})(1-q_{3})}{2(s_{2}-s_{1})}+1,\\
q_{2}=\frac{\Phi_{min}^{2s_{2}-2s_{3}}(s_{1}-s_{3})(1-q_{3})}{(s_{1}-s_{2})}+1.
\end{align}

\begin{figure}
\centering
\includegraphics[scale=0.35]{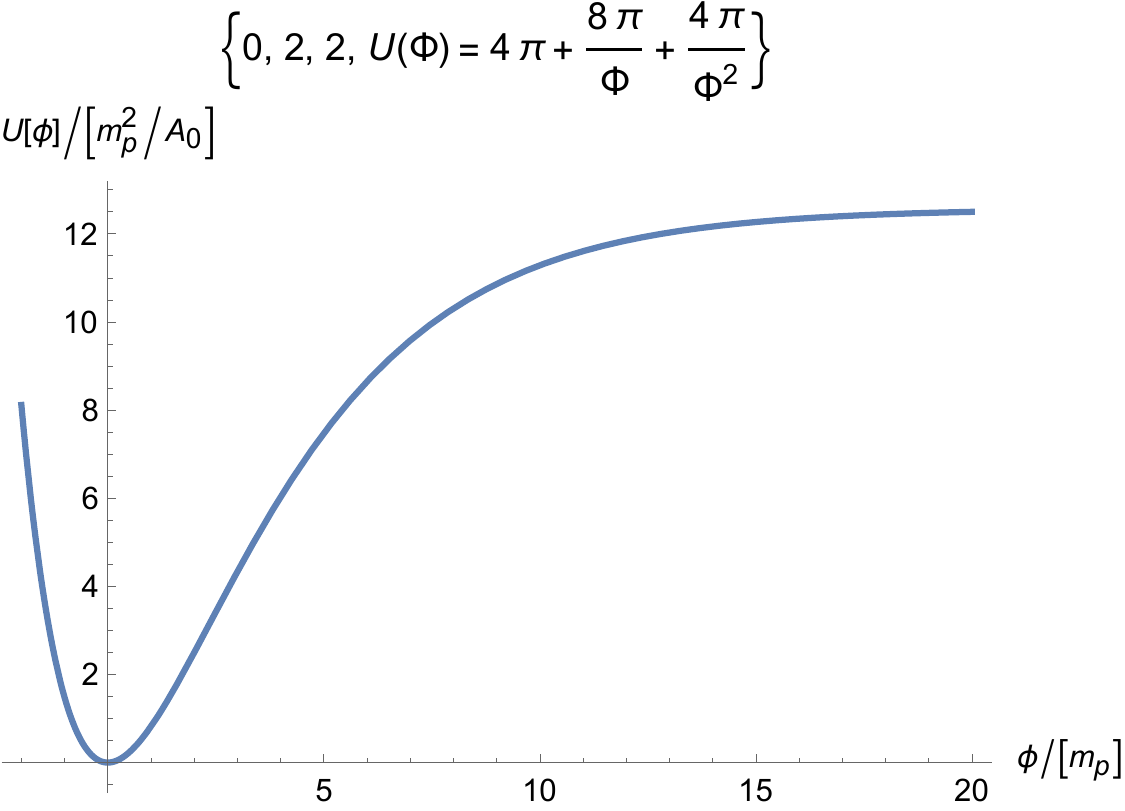}
\includegraphics[scale=0.35]{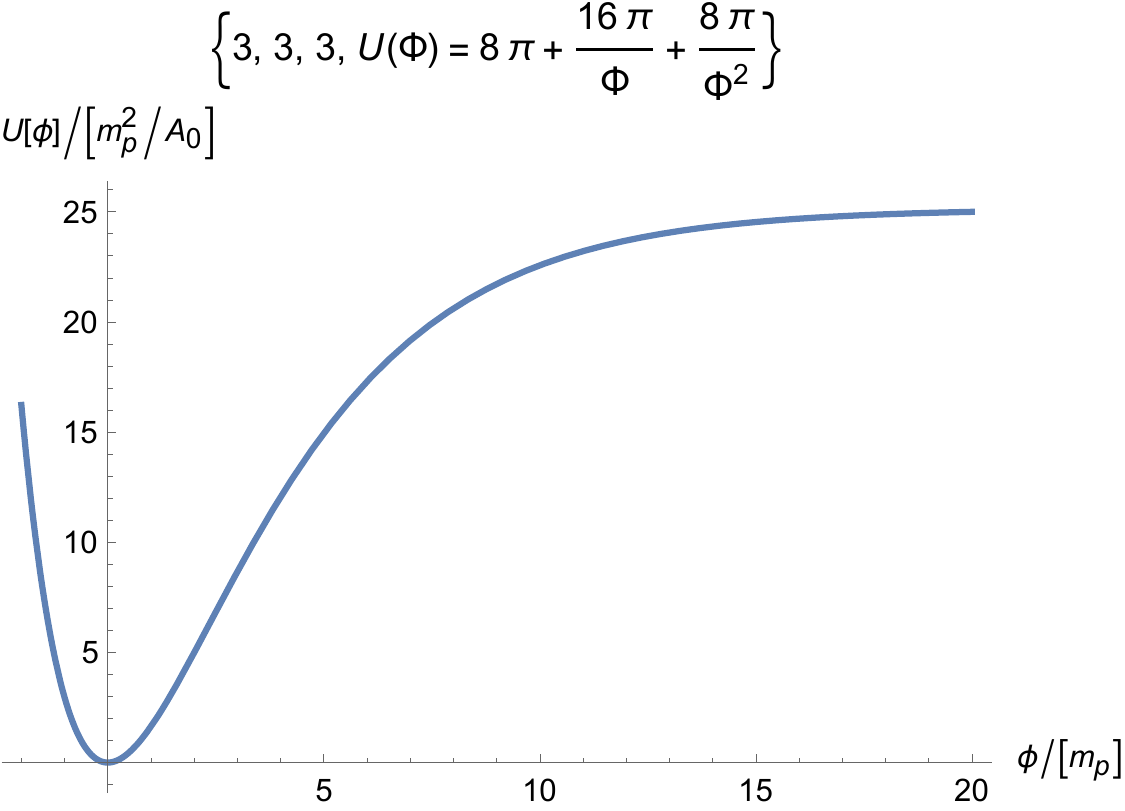}

\includegraphics[scale=0.35]{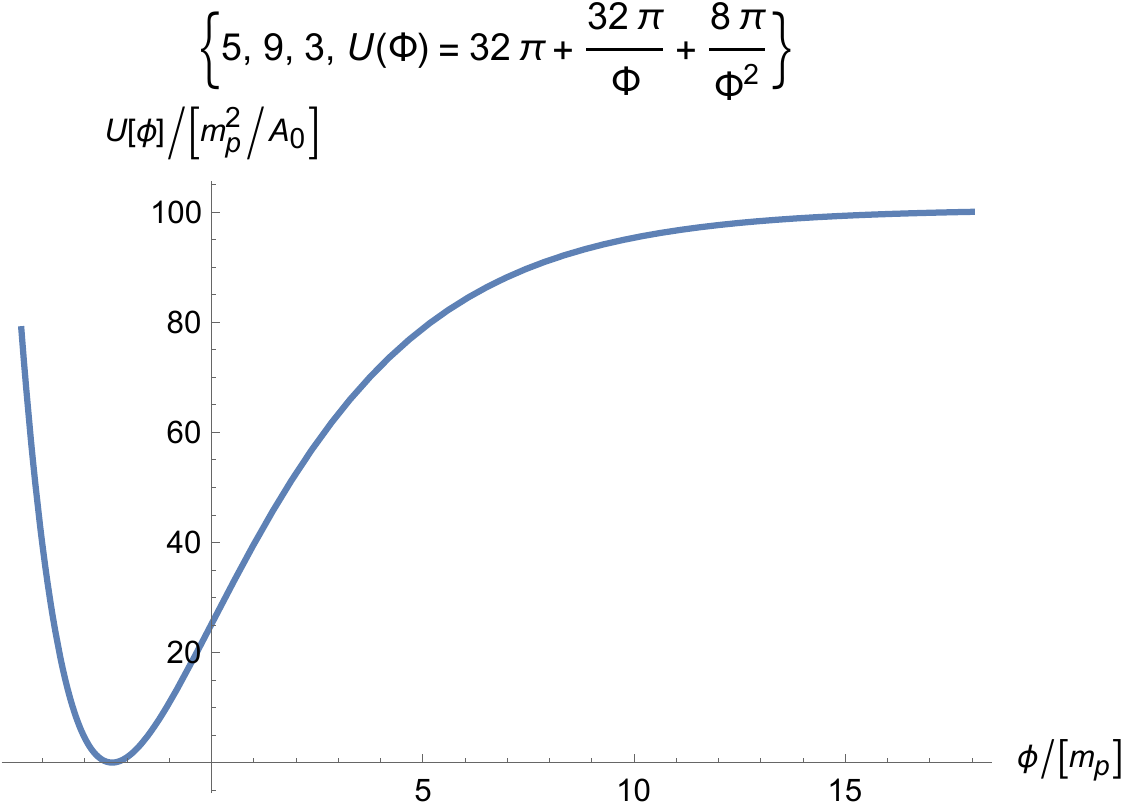}
\includegraphics[scale=0.35]{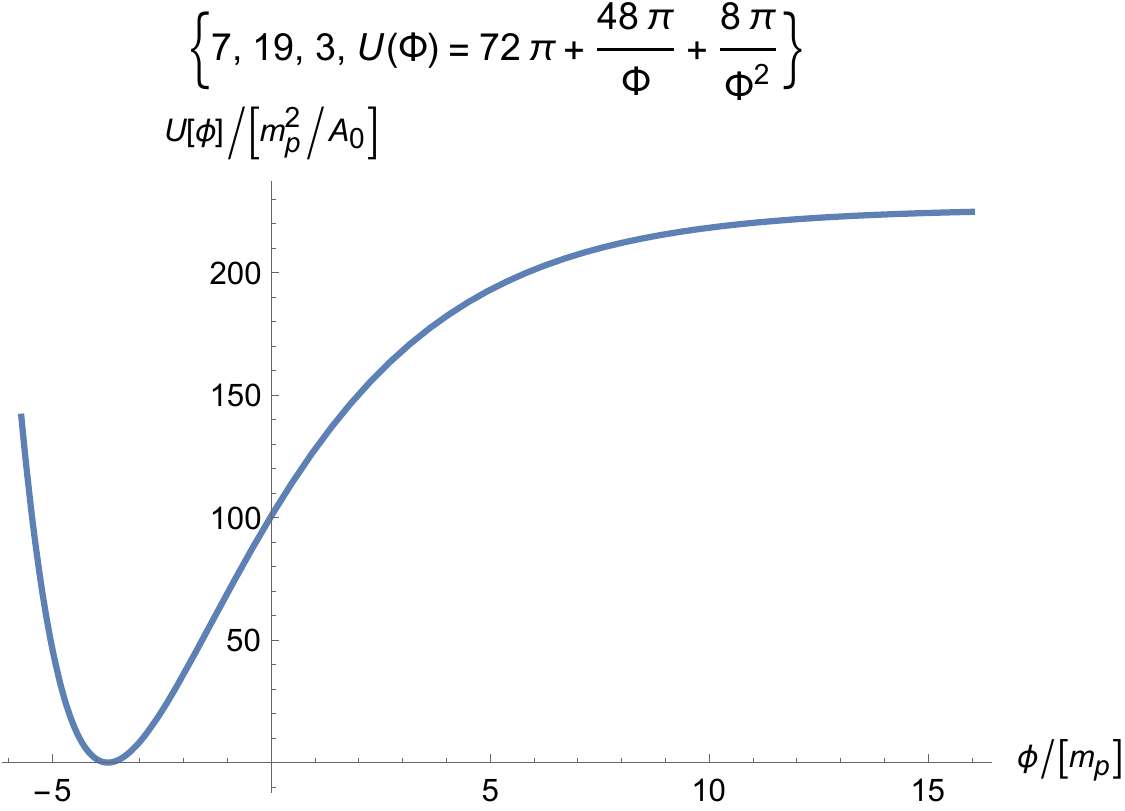}
\caption{Pictures for inflation potential. The first figure depicts the potential of the special case given by equation \eqref{EHaction1}, where $(q_{1},q_{2},q_{3})=(0,2,2)$. The subsequent figures illustrate potentials $U(\Phi)$ in equation \eqref{potential0} for varying parameter sets of $(q_{1},q_{2},q_{3},n)$. Specifically, the second picture in the first row corresponds to the parameter set $(q_{1},q_{2},q_{3},n)=(3,3,3,1)$. In the second row, the first picture represents the potential for $(q_{1},q_{2},q_{3},n)=(5,9,3,2)$, while the last picture corresponds to $(q_{1},q_{2},q_{3},n)=(7,19,3,3)$.}
\label{fig3}
\end{figure}

To ensure the genus of the three tori $q_{1},q_{2},q_{3}\in\mathbb{N}$, we can set $\Phi_{min}=-\frac{1}{n},s_{1}=\frac{1}{2},s_{2}=0$, then
\begin{align}
q_{1}&=1+n^{2s_{3}-1}(q_{3}-1)s_{3},\\
q_{2}&=n^{2s3}\left[1+n^{-2s_{3}}+2s_{3}\left(q_{3}-1\right)-q_{3}\right],
\end{align}
where  $q_{3}\geq2,s_{3}>0,s_{3}\in\mathbb{Z}$. In this case
\begin{align}
U(\Phi)=\frac{4\pi m_{p}^{2}\text{\ensuremath{\left(q_{3}-1\right)}}\Phi^{-2s_{3}-1}}{A_{0}n}\left[n\Phi+\left(n\Phi\right)^{2s_{3}}\left(2s_{3}+\left(2s_{3}-1\right)n\Phi\right)\right],\label{potential0}
\end{align}
with $\Phi=-e^{\sqrt{\lambda^{-1}}\phi/m_{p}}$ and $n,q_{3}\in\mathbb{N}$. The pictures of this potential are shown in Figure \ref{fig3}. It can be seen from the listed potential functions that the lowest point of potential energy is located at $U(\Phi_{min})=0$. When $\phi\gg 1$, the potential energy tends to be a flat potential, which has the characteristics of the potential function of the Starobinsky model. An important property of the potential $U(\Phi)$ is that the spectral index $n_{s}$ and tensor-to-scalar ratio $r$ are independent of the factor $(q_{3}-1)$ and parameter $n$. Therefore, the only constraint imposed on the third torus is $q_{3}\geq 2$. For more details, please refer to the  \ref{Appendix}. Moreover, we rediscover the special case in Subsection \ref{subectionA} by choosing $\Phi_{min}=1,s_{1}=\frac{1}{2},s_{2}=0,s_{3}=1,q_{1}=0,q_{2}=q_{3}=2$.

\section{Cosmological Inflation}\label{sectionIII}
In this section, we aim to numerically compute the inflation parameters corresponding to the inflation models discussed in Section \ref{sectionII}. The inflaton $\phi$ is clearly associated with the size function $\Phi(x^{\mu})$
of the extra dimensions. Therefore, the vibrations of the inflaton $\phi$ can be considered as the conformal vibrations of the extra dimensions.

The slow-roll parameters $\epsilon$ and $\eta$, along with the spectral index $n_{s}$ and tensor-to-scalar ratio $r$, are determined as follows:
\begin{align}
\epsilon(\phi) & =\frac{m_{p}^{2}}{2}\left(\frac{U_{\phi}}{U}\right)^{2},\\
\eta(\phi) & =m_{p}^{2}\left(\frac{U_{\phi\phi}}{U}\right),\\
n_{s}&=1-6\epsilon(\phi_{i})+2\eta(\phi_{i}),\\
r&=16\epsilon(\phi_{i}).
\end{align}

\begin{figure}
\centering
\includegraphics[scale=0.8]{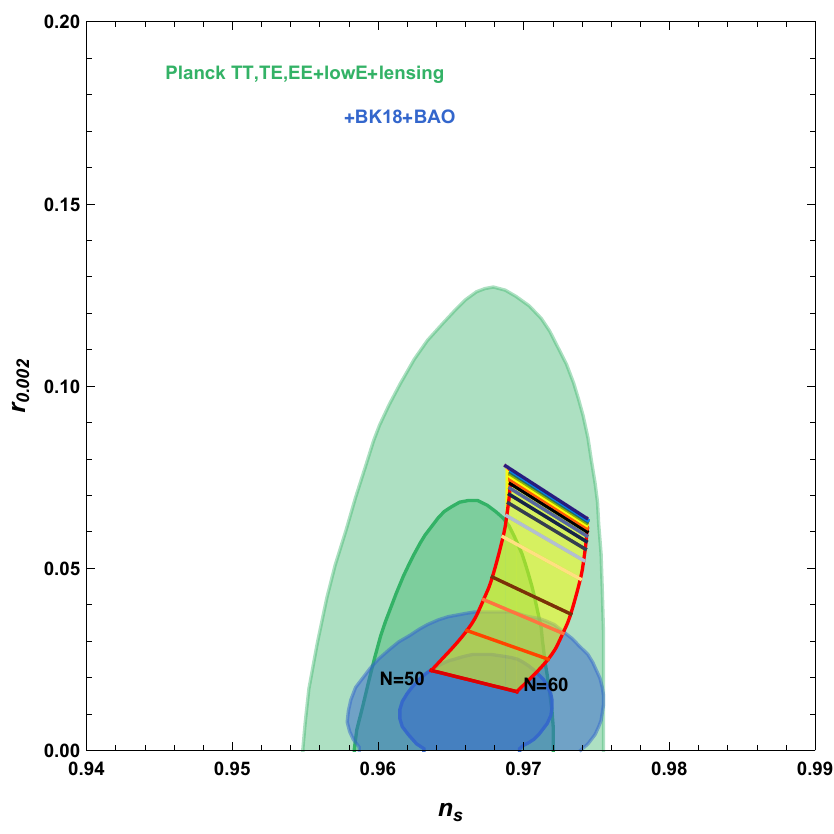}
\caption{The $n_{s}-r$ constraints for potential \eqref{potential0}. The green contours are from Planck 2018 data \cite{Planck:2018jri}. The blue contours are from 2021 BICEP/Keck data \cite{BICEP:2021xfz}. Here, we choose $s_3=1,2,3,4,7,10,...,37,40$. The straight lines of varying colors represent distinct values of $s_{3}$. The lowest red straight line represents $s_{3}=1$. As $s_{3}$ increases, the corresponding $n_{s}-r$ line progressively shifts beyond the blue observational region (See the yellow region.). However, as $s_{3}$ continues to grow, these lines tend to cluster around the purple line at the top.}
\label{fig1}
\end{figure}
The e-folding number is
\begin{align}
N=-\frac{1}{m_{p}^{2}}\int_{\phi_{i}}^{\phi_{e}}\frac{U}{U_{\phi}}d\phi,
\end{align}
where $\phi_{i}$ represents the initial value of the inflaton and $\phi_{e}$ represents its end value. The numerical results of the general case with potential  \eqref{potential0} are shown in Figure \ref{fig1}. The lowest three lines in Figure \ref{fig1} correspond to the cases of $s_3=1,2,3$, respectively. They are consistent with the stronger $n_s-r$ constraints by Planck TT,TE,EE+lowE+lensing+BK18+BAO (blue region). With the increasing $s_3$, the convergence of $n_s-r$ happens inside the green region of weak $n_s-r$ constraints. As a result, the inflation models from the compactification methods $\mathcal{M}^{3,1}\times T_{q_{1}}^{2}\times T_{q_{2}}^{2}\times T_{q_{3}}^{2}\times S^{1}$ can be in good accordance with the latest observations. The optimal value for $s_{3}$ is $s_{3}=1$, as its corresponding $n_{s}-r$ line approaches the center of the observational constraints, which is the same as the result of the special case discussed in Subsection \ref{subectionA}.

For the special case \eqref{EHaction1}, the inflation exits as $\epsilon(\phi_{e})=1, \phi_{e}=1.182m_{p}$. For e-folding $N=60$, the solutions are $(\phi_{i},\epsilon,\eta,n_{s},r)=(8.986m_{p},0.0010,-0.012,0.9695,0.016)$. For e-folding $N=50$, $(\phi_{i},\epsilon,\eta,n_{s},r)=(8.503m_{p},0.0013,-0.014,0.9637,0.022)$. These results match the Planck 2018 data \cite{Planck:2018jri}. Taking into account the observational constraint imposed on the power spectrum \cite{Planck:2018vyg}
\begin{align}
P_{s} =\frac{U(\phi_{i})}{24\pi^{2}m_{p}^{4}\epsilon(\phi_{i})}\approx2.1\times10^{-9}.
\end{align}
When e-folding $N=60$, the standard areas of the three extra-dimensional tori can be obtained as follows:
\begin{align}
A_{0}=\frac{4\pi}{\zeta}l_{p}^{2}=1.43\times10^{-58}[\text{m}]^{2},
\end{align}
where $\zeta=5.79\times10^{-10}$ is coupling constant of potential
\begin{align}
U(\phi)=\zeta m^{4}_{p}\left(1-e^{-\sqrt{\frac{3}{23}}\frac{\phi}{m_{p}}}\right)^{2},
\end{align}
and $l_{p}$ is the reduced Planck length. The symbol $[\text{m}]$ is used to represent the unit of meters. To approximate the size of the extra dimensions, we can assume that the shape of the extra dimensions $T_{q_{1}}^{2}$ is roughly spherical. Consequently, the radius of this sphere would be
\begin{equation}
\mathcal{R}=\sqrt{\frac{A_{0}}{8\pi}}=29387l_{p}=2.38\times10^{-30}[\text{m}].\label{extrascale}
\end{equation}

The range of the scalar field $\phi\in(\phi_{i},\phi_{e}),\phi_{i}\in(8.5025m_{p},8.986m_{p})$ determines the range of the scale function of the extra dimensions during inflation, as specified below:
\begin{align}
1.417 & \leq\Phi\leq12.271\ (\text{for }N=50)\\
1.417 & \leq\Phi\leq14.153\ (\text{for }N=60)
\end{align}

\begin{figure}
\centering
\includegraphics[scale=0.45]{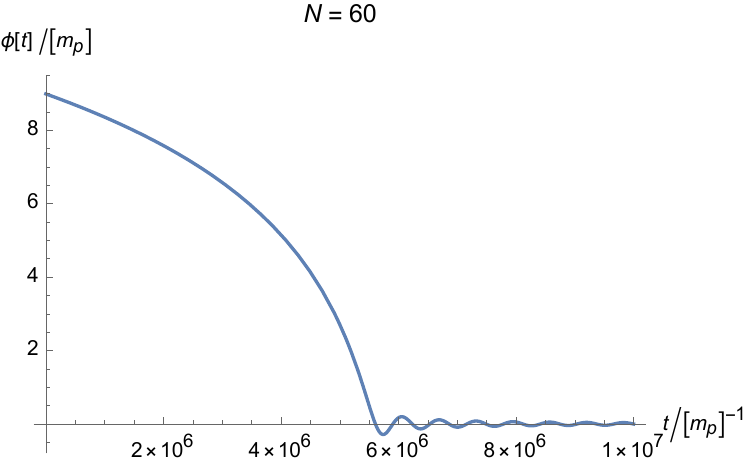}
\includegraphics[scale=0.45]{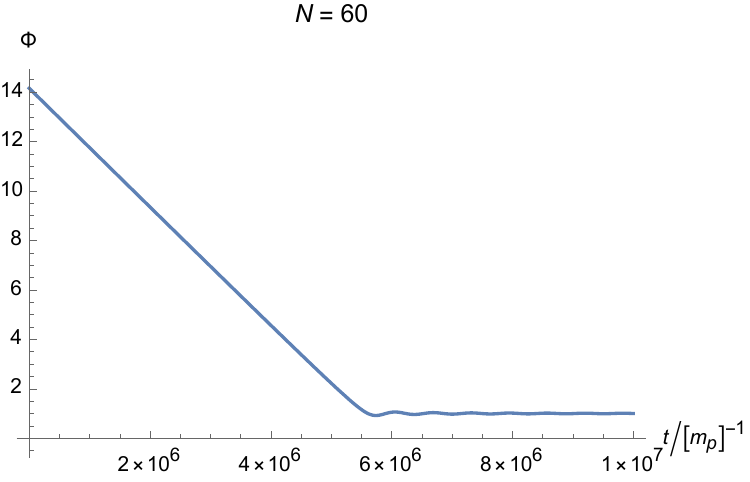}
\caption{The dynamics of the inflaton and scale function for e-folding $N=60$. The first figure depicts the scalar field rolling down to 0 at $N=60$, while the second figure shows the scale function converging to 1. During inflation, the dynamics of the scale function $\Phi$ exhibit a decreasing trend.}
\label{fig2}
\end{figure}

It shows that the scale of the extra dimensions is considerably small at order $O(10^{5}l_{p})$. In our compactification methods, the eleventh dimension is compactified into a circle, indicating that the theory resulting from compactification is Type IIA string theory in 10 dimensions. Subsequently, the string length $l_{s}$ relates to the radius of the eleventh dimension $\mathcal{R}_{10}$ as follows \cite{Ydri:2018sjn}:
\begin{align}
l_{s}=\frac{l_{p(11)}}{\sqrt{\frac{\mathcal{R}_{10}}{l_{p(11)}}}}.\label{stringlength}
\end{align}
In today's universe, where $\Phi=1$, we postulate that the string length is represented by $l_{s}=\xi l_{p(11)}$ and its corresponding eleventh dimension radius is given by $\widetilde{\mathcal{R}}_{10}=\xi^{-2}l_{p(11)}$. The 11D Newtonian gravitational constant $G_{N}^{(11)}$ and the 4D Newtonian gravitational constant $G$ are related through the volume of the extra dimensions \eqref{massrelation} as follows:
\begin{align}
G^{(11)}_{N}&=\pi \xi^{-2}l_{p(11)}A_{0}^{3}G.
\end{align}
Consequently, the relationship between the 4D reduced Planck length $l_{p}$ and the 11D reduced Planck length $l_{p(11)}$ is expressed by equations
\begin{align}
l_{p}&=\frac{l_{p(11)}^{\frac{9}{2}}}{\left(\pi\xi^{-2}l_{p(11)}A_{0}^{3}\right)^{\frac{1}{2}}},\\
l_{p(11)}&=\left(\pi A_{0}^{3}\xi^{-2}\right)^{\frac{1}{8}}l_{p}^{\frac{1}{4}}=l_{p}\left[\frac{\left(4\pi\right)^{4}}{4\zeta^{3}\xi^{2}}\right]^{\frac{1}{8}}=8676.7\xi^{-\frac{1}{4}}l_{p}.
\end{align}
However, during the inflationary period, the eleventh dimension experiences a scaling factor of $\Phi^{-3}$. Hence, for e-folding $N=60$, the radius adjusts to $\mathcal{R}_{10}=\xi^{-2}\Phi^{-3}l_{p(11)}\in\xi^{-2}[0.000352l_{p(11)},0.351l_{p(11)}]$.  The string length during this period can be determined using equation \eqref{stringlength} as follows:
\begin{align}
l_{s}\in\xi\left[1.7l_{p(11)},53.2l_{p(11)}\right].
\end{align}
The standard tori extra dimensions have a size scale of $\mathcal{R}=29387l_{p}=3.39\xi^{\frac{1}{4}}l_{p(11)}$. Assuming the loop-order quantum corrections of eleven-dimensional supergravity \cite{Hyakutake:2006aq} can be neglected due to their small coupling constant $\alpha'\sim O(l^{2}_{s})$, the string length and the size scale of the tori must satisfy the condition $\frac{l_{s}}{\mathcal{R}}\ll1$. This limitation imposes a constraint on the parameter $\xi\ll0.025$. Therefore, we anticipate that the current string length meets the criterion $l_{s}=8676.7\xi^{\frac{3}{4}} l_{p}\ll 551.9l_{p}$.

The dynamics of the inflaton and the scale function are shown in figure \ref{fig2}. It demonstrates that the three tori contract conformally, while the eleventh dimension expands. The scale function oscillates around the point where $\Phi(x^{\mu})$ is equal to 1 after the process of inflation, see \ref{Appendix3} for a brief discussion on exit inflation.

\section{Dark Sector from Extra Dimensions}\label{quantumEFF}
In this section, we concentrate on the quantum effects from the inflaton in the four-dimensional extended spacetime specifically for the case presented in Subsection \ref{subectionA}. As shown in Equation \eqref{extrascale}, the extra dimensions exist at the sub-Planck scale $O(10^{5}l_{p})$, and their conformal vibrations can be interpreted as the vibrations of the inflaton $\phi$. Referring to Figure \ref{fig2}, it becomes evident that the inflaton field arrives at a significantly small value, namely $\phi\ll m_{p}$, in today's universe. Hence, the potential given in \eqref{specialpotential} can be expanded up to the second order as
\begin{align}
V(\phi)\approx \frac{1}{2}m^{2}_{\phi}\phi^{2},
\end{align}
where
\begin{align}
m_{\phi}=\sqrt{\frac{4\zeta}{23}}m_{p}=10^{-5}m_{p}=2.4\times10^{10}\frac{\text{TeV}}{\text{c}^{2}}.
\end{align}
The rest mass of the inflaton is so heavy that detecting it exceeds the capabilities of current accelerators. The energy scale at which the inflaton appears is approximately $\Lambda_{\phi}=m_{\phi}c^{2}\approx 2.4\times10^{10}\text{TeV}$. This energy scale is significantly higher than those achievable by the Large Hadron Collider (LHC), making it challenging to directly observe the inflaton in accelerators. However, gravity is a long-range force, and Newton's law with the modification by the inflaton can be verified through astronomical observations.

Following the second quantization method, the inflaton field emerges as a massive scalar quantum field. The tree-level propagator for graviton with momentum $k^{\mu}=(k_{0},\mathbf{k})$ is given by \cite{Mohanty:2022abo}
\begin{align}
\Pi_{\mu\nu\rho\sigma}(x-x')&=\int\frac{dk_{0}}{2\pi}\frac{d^{3}k}{(2\pi)^{3}}\frac{-iP_{\mu\nu\rho\sigma}}{k_{0}^{2}-|\mathbf{k}|^{2}+i\varepsilon}e^{-ik_{0}(t-t')}e^{i\mathbf{k}\cdot(\mathbf{x}-\mathbf{x}')},
\end{align}
where
\begin{align}
P_{\mu\nu\rho\sigma}&=\frac{1}{2}\left(\eta_{\mu\rho}\eta_{\nu\sigma}+\eta_{\mu\sigma}\eta_{\nu\rho}-\eta_{\mu\nu}\eta_{\rho\sigma}\right).
\end{align}
The metric is expanded to first order using a perturbation tensor $h_{\mu\nu}$ as follows:
\begin{align}
g^{\mu\nu}=\eta^{\mu\nu}+\frac{2}{m_{p}}h^{\mu\nu},
\end{align}
where the Minkowski metric $\eta_{\mu\nu}=\text{diag}(-1,1,1,1)$ in the local reference frames. Consider the action \eqref{EHaction1}  along with the matter term $\mathcal{L}_{\text{matter}}$, the energy stress tensor can be represented as
\begin{align}
\mathcal{T}^{\mu\nu}=\mathcal{T}^{\mu\nu}_{\phi}+\mathcal{T}^{\mu\nu}_{\text{(matter)}},
\end{align}
where the energy stress tensor of the inflaton $\mathcal{T}^{\mu\nu}_{\phi}$, as well as the density $\rho$, are defined by
\begin{align}
\mathcal{T}^{\mu\nu}_{\phi}&=\partial^{\mu}\phi\partial^{\nu}\phi-\frac{1}{2}\eta^{\mu\nu}\partial^{\alpha}\phi\partial_{\alpha}\phi-\eta^{\mu\nu}V(\phi),\\
\rho&=\frac{1}{2}\dot{\phi}^{2}+\frac{1}{2}\partial_{i}\phi\partial^{i}\phi+V(\phi).
\end{align}
In classical mechanics, when considering two massive point particles with masses $M_{1}$ and $M_{2}$, Newton's gravity law has modification due to the interactions from the inflaton. The four-point tree-level scattering amplitude for the matter-graviton-matter process can be expressed as the effective action of the second order \cite{Mohanty:2022abo}
\begin{align}
iS^{(2)}&=\frac{1}{m_{p}^{2}}\int d^{4}xd^{4}x'\mathcal{T}_{1}^{\mu\nu}(x)\Pi_{\mu\nu\rho\sigma}(x-x')\mathcal{T}_{2}^{\rho\sigma}(x'),\label{amplitude}
\end{align}
where
\begin{align}
\mathcal{T}^{\mu\nu}_{\text{(matter1)}}&=M_{1}u_{1}^{\mu}u_{1}^{\nu}\delta^{3}(\mathbf{x}-\mathbf{q}_{1}(t)),\\
\mathcal{T}^{\mu\nu}_{\text{(matter2)}}&=M_{2}u_{2}^{\mu}u_{2}^{\nu}\delta^{3}(\mathbf{x}'-\mathbf{q}_{2}(t)).
\end{align}
The four-vector $u^{\mu}=(1-\mathbf{v}^{2})^{-\frac{1}{2}}(1,\mathbf{v})=I^{\mu}+O(\mathbf{v})$, where $I^{\mu}=(1,0,0,0)$. This four-point tree-level scattering amplitude contains both the Newton potential part between $M_{1}$ and $M_{2}$, denoted as \cite{Mohanty:2022abo}
\begin{align}
iS^{(2)}_{\text{(matter)}}=\frac{1}{m_{p}^{2}}\int d^{4}xd^{4}x'\mathcal{T}_{\text{(matter1)}}^{\mu\nu}(x)\Pi_{\mu\nu\rho\sigma}(x-x')\mathcal{T}_{\text{(matter2)}}^{\rho\sigma}(x'),
\end{align}
 and the self-gravity interaction term of inflaton, denoted as
\begin{align}
iS^{(2)}_{\phi}=\frac{1}{m_{p}^{2}}\int d^{4}xd^{4}x'\mathcal{T}^{\mu\nu}_{\phi}(x)\Pi_{\mu\nu\rho\sigma}(x-x')\mathcal{T}^{\rho\sigma}_{\phi}(x'),
\end{align}
The remaining two terms in \eqref{amplitude} are the modification for Newton's gravity law between $M_{1}$ and $M_{2}$
\begin{equation}
\begin{aligned}
i\left(S^{(2)}-S^{(2)}_{\text{(matter)}}-S^{(2)}_{\phi}\right)=&i\frac{M_{1}}{m_{p}^{2}}\int dt\int\frac{d^{3}\mathbf{k}}{(2\pi)^{3}}\frac{1}{|\mathbf{k}|^{2}-i\varepsilon}\int d^{3}\mathbf{x}F(t,\mathbf{x})e^{i\mathbf{k}\cdot(\mathbf{q}_{1}(t)-\mathbf{x})}\\
&+i\frac{M_{2}}{m_{p}^{2}}\int dt\int\frac{d^{3}\mathbf{k}}{(2\pi)^{3}}\frac{1}{|\mathbf{k}|^{2}-i\varepsilon}\int d^{3}\mathbf{x}'F(t,\mathbf{x}')e^{i\mathbf{k}\cdot(\mathbf{x}'-\mathbf{q}_{2}(t))}\\
&+O(\mathbf{v}_{1},\mathbf{v}_{2}),
\end{aligned}
\end{equation}
where
\begin{align}
F(t,\mathbf{x})=\left(\frac{\partial\phi(t,\mathbf{x})}{\partial t}\right)^{2}-V(\phi(t,\mathbf{x})),\label{fourier2}
\end{align}
and we have taken $k_{0}\sim\frac{v}{r}\rightarrow0$ for the massless graviton in $\mathbf{v}\rightarrow0$ limit. The modified Newton potential is now given by
\begin{equation}
\begin{aligned}
W(\mathbf{q}_{1}(t),\mathbf{q}_{2}(t))=&-\frac{M_{1}M_{2}}{8\pi m_{p}^{2}}\frac{1}{\left|\mathbf{q}_{1}(t)-\mathbf{q}_{2}(t)\right|}-\frac{M_{1}}{m_{p}^{2}}\int\frac{d^{3}\mathbf{k}}{(2\pi)^{3}}\frac{1}{|\mathbf{k}|^{2}-i\varepsilon}\int d^{3}\mathbf{x}F(t,\mathbf{x})e^{i\mathbf{k}\cdot(\mathbf{q}_{1}(t)-\mathbf{x})}\\
&-\frac{M_{2}}{m_{p}^{2}}\int\frac{d^{3}\mathbf{k}}{(2\pi)^{3}}\frac{1}{|\mathbf{k}|^{2}-i\varepsilon}\int d^{3}\mathbf{x}'F(t,\mathbf{x}')e^{i\mathbf{k}\cdot(\mathbf{x}'-\mathbf{q}_{2}(t))}+O(\mathbf{v}_{1},\mathbf{v}_{2}).\label{fourier00}
\end{aligned}
\end{equation}
After doing the calculation on the integrals with respect to momentum $\mathbf{k}$, the modified Newton potential can be expressed as
\begin{equation}
\begin{aligned}
W(\mathbf{q}_{1}(t),\mathbf{q}_{2}(t))=&-\frac{M_{1}M_{2}}{8\pi m_{p}^{2}}\frac{1}{\left|\mathbf{q}_{1}(t)-\mathbf{q}_{2}(t)\right|}-\frac{M_{2}}{4\pi m_{p}^{2}}\int d^{3}\mathbf{x} \frac{F(t,\mathbf{x})}{\left|\mathbf{q}_{2}(t)-\mathbf{x}\right|}\\
&-\frac{M_{1}}{4\pi m_{p}^{2}}\int d^{3}\mathbf{x}' \frac{F(t,\mathbf{x}')}{\left|\mathbf{q}_{1}(t)-\mathbf{x}'\right|},\label{modifiedpotential1}
\end{aligned}
\end{equation}
The equation of motion for inflaton $\phi(t,\mathbf{x})$ is given by $\nabla_{\mu}T^{\mu\nu}_{\phi}(x)=0$ in the FRW metric, which is
\begin{align}
\ddot{\phi}(t,\mathbf{x})+3H\dot{\phi}(t,\mathbf{x})-\frac{1}{a^{2}(t)}\partial_{i}\partial^{i}\phi(t,\mathbf{x})+m_{\phi}^{2}\phi(t,\mathbf{x})=0,\label{eomphi}
\end{align}
where $H=\frac{\dot{a}(t)}{a(t)}$ is the Hubble constant. As we know, the current universe enters a de Sitter phase since the dark energy dominates the cosmic evolution, therefore $H$ is almost a constant. When the inflaton field $\phi(t,\mathbf{x})$ has spherical symmetry, there is $\partial_{i}\partial^{i}\phi(t,r)=\frac{1}{r^{2}}\partial_{r}(r^{2}\partial_{r}\phi(t,r))$. Solving the equation of motion \eqref{eomphi} yields the expression for $F(t,\mathbf{x})$, where the FRW metric goes to the Minkowski spacetime as $a(t)\simeq1$.

\subsection{Dark Matter}\label{darkmatter}
We consider the solution in time-splitting form $\phi(t,\mathbf{x})=\phi_{1}(t)\phi_{2}(\mathbf{x})$. Assuming that the spatial part $\phi_{2}(\mathbf{x})$ has spherical symmetry, we can deduce the following equations:
\begin{align}
\ddot{\phi}_{1}(t)+3H\dot{\phi}_{1}(t)+m^{2}_{\phi}\phi_{1}(t)&=0,\label{eomphi1}\\
\frac{1}{r^{2}}\partial_{r}(r^{2}\partial_{r}\phi_{2}(r))&=0.\label{eomphi2}
\end{align}
Note that the Hubble constant is approximately $H\approx 6.1\times10^{-61}m_{p}$. Therefore, there always exists a relation $m_{\phi}=10^{-5}m_{p}\gg H$. In the limit $m_{\phi}\gg H$ and $a(t)\approx 1$ for today's universe, the solutions to the equations of motion, given by \eqref{eomphi2} and \eqref{eomphi1}, can be described by
\begin{align}
\phi_{1}(t)&\approx 2\phi_0 a^{-\frac{3}{2}}\cos(m_{\phi} t),\\
\phi_{2}(r)&=C_{1}+\frac{C_{2}}{r},
\end{align}
where $\phi_{0},C_{1},C_{2}$ are constants and $m_{\phi}=\sqrt{\frac{4\zeta}{23}}$ in Planck unit. The time average of the function $F(t,\mathbf{x})$ can now be reformulated as
\begin{align}
\left\langle F(t,\mathbf{x})\right\rangle_{t}=\frac{m_{\phi}}{2\pi}\int_{0}^{\frac{2\pi}{m_{\phi}}}\left(\dot{\phi}^{2}(t,r)
-\frac{1}{2}m_{\phi}^{2}\phi^{2}(t,r)\right)dt=\phi^{2}_{2}(r)a^{-3}m^{2}_{\phi}\phi_0^2.
\end{align}
The function $F(t,\mathbf{x})$ with a background evolution at the time-averaged level is given by
\begin{align}
\left\langle F(t,\mathbf{x})\right\rangle_{t}=\frac{1}{2}\phi^{2}_{2}(r)\overline{\rho_{DM}(t)}\,,\label{density}
\end{align}
where $\overline{\rho_{DM}(t)}=2a^{-3}m^{2}_{\phi}\phi_0^2=3m_{p}^{2}H^{2}\times\Omega_{DM}$, here $\Omega_{DM}$ denotes the fraction parameter of dark matter. The time evolution of $\overline{\rho_{DM}(t)}$ can tell the coldness of the dark matter, i.e. its equation of state $w_{DM}\simeq0$, which has been already accepted in the standard cosmological model.

Now, setting $\mathbf{q}_{1}(t)=0$ and locating the center of the dark matter distribution at the origin $\mathbf{q}_{1}(t)$, we follow the method in Ref. \cite{Fitzpatrick_2012} to calculate Eq. \eqref{modifiedpotential1}. Hence Newton's law of gravity for $M_{2}$ is given as follows
\begin{align}
F_{2}(R)=-G\frac{M_{1}M_{2}}{R^{2}}-4\pi GM_{2}\overline{\rho_{DM}(t)}\left(\frac{ C_{1}^{2}}{3}R+\frac{C^{2}_{2}}{R}+C_{1}C_{2}\right),\label{gravity1}
\end{align}
where we define $R=\left|\mathbf{q}_{2}(t)\right|$. Here $F_{2}(R)$ is the gravitational force exerted on $M_2$ and $G$ is the Newtonian gravitational constant. The modified portion of Newton's gravity law describes a spherically symmetric distribution of the inflaton, which can be interpreted as a form of dark matter located around $M_{1}$. The constants $C_{1}, C_{2}$ can be determined through astronomical observations at the galactic scale. If $C_1=0$, the flat rotation curves can be realized because the speed of object $M_2$ becomes $v_2^2 \equiv -F_2R/M_2\simeq GK$ at a very large $R$ with $K=4\pi C_2^{2}\overline{\rho_{DM}(t)}$.

Note that we solely focus on the leading order contribution in the limit $m_{\phi}\gg H$. Assuming $m_{\phi}^{2}=m^{2}_{1}+m^{2}_{2}$,  and considering the limits $m_{1}\gg H$ and $m_{1}\gg m_{2}$ for both the leading and sub-leading contributions, the equations for $\phi_{1}(t)$ and $\phi_{2}(r)$ can be described by equations
\begin{align}
\ddot{\phi_{1}}(t)+3H\dot{\phi_{1}}(t)+m_{1}^{2}\phi_{1}(t)&=0,\\
-\frac{1}{r^{2}}\partial_{r}\left(r^{2}\frac{\partial\phi_{2}(r)}{\partial r}\right)+a^{2}m_{2}^{2}\phi_{2}(r)&=0.
\end{align}
The solutions to these two equations are provided in
\begin{align}
\phi_{1}(t)&\approx 2\phi_0a^{-\frac{3}{2}}\cos(m_{1} t),\\
\phi_{2}(r)&=C\frac{e^{-am_{2}r}}{r}+D\frac{e^{am_{2}r}}{r},\label{solution2}
\end{align}
where $\phi_{0}, C$ is constant and $D=0$ since we should cancel the blow-up term containing $e^{am_{2}r}$ when $r$ goes to infinity. Under these circumstances, the function $F(t,\mathbf{x})$ of the inflaton retains the form given in \eqref{density}, but with $m_{\phi}$ replaced by $m_{1}$ and $\phi_{2}(r)$ as specified in \eqref{solution2}. Therefore, the modified Newton potential is still given by \eqref{modifiedpotential1} and the corresponding Newton's gravity law for $M_{2}$ is given by
\begin{align}
F_{2}(R)=-\frac{GM_{1}M_{2}}{R^{2}}-\frac{2\pi GM_{2}\overline{\rho_{DM}(t)}C^{2}\left(1-e^{-2am_{2}R}\right)}{a m_{2}R^{2}}.\label{gravity2}
\end{align}
The characteristic length of the Yukawa term, given by $\lambda=\frac{\hbar}{2am_{2}c}$, constrains the value of $m_{2}$ based on experimental results on Yukawa gravitational potential \cite{Tan:2020vpf,Lee:2020zjt} and observation results on galaxies. Considering $m_2$ correction, when $m_2R\gg1$, the velocity decays in the manner of $v_2^{2}\propto R^{-1}$ which reflects the original Newton's gravity law still holds at an extremely large distance but additional mass from dark matter to $M_1$ should be included as the central gravitational source. When $m_{2}R\ll 1$, Eq. \eqref{gravity1} in the case of $C_1=0$ is recovered, and the velocity of $M_{2}$ is directly expressed by equation
\begin{align}
v_2^{2}\approx\frac{GM_{1}}{R}+GK,\label{velocity}\\
K=4\pi C^{2}\overline{\rho_{DM}(t)},
\end{align}
where $R$ can be determined using Newtonian acceleration $a_{N}$
\begin{align}
R=\sqrt{\frac{GM_{1}}{a_{N}}}.
\end{align}
Additionally, the acceleration of $M_{2}$ is provided by
\begin{align}
a_{2}=\frac{v_{2}^{2}}{R}=a_{N}\left(1+\sqrt{\frac{a_{0}}{a_{N}}}\right),
\label{mond}
\end{align}
where the characteristic acceleration is defined as
\begin{align}
a_{0}=\frac{K^{2}G}{M_{1}}.\label{a0}
\end{align}
Equation \eqref{mond} recovers the formulation of the MOND theory \cite{Milgrom:1983zz,Milgrom:1983ca,Bekenstein:1984tv,Milgrom:1986ib}. Then, we find the Tully-Fisher relation from \eqref{velocity}\eqref{a0} as $R\rightarrow \infty$
\begin{align}
v_{2}^{4}=GM_{1}a_{0}.\label{v4}
\end{align}
These are two of the ways in which extra dimensions affect Newton's gravity law. In this scenario, the extra dimensions are viewed as a spherically symmetric distribution of dark matter. Astronomical observations, as referenced in \cite{Das:2023zvp}, reveal that the characteristic acceleration is of the order of $a_{0}\sim 10^{-8} [\text{cm/}\text{s}^{2}]$ and the radii of galaxies are approximately  $R_{\text{max}}\sim10^{2}[\text{kpc}]\approx 3.8\times10^{55}l_{p}$.  Consequently, the characteristic mass becomes $m_{2c}=1/R_{\text{max}}\approx 2.6\times10^{-56}m_{p}$. Therefore, we observe the relations $m_{1}\gg m_{2c}\gg m_{2}$. The modified portion of Newton's gravity law is associated with both the mass of the inflaton $m_{\phi}$ and the strength of the inflaton field
\begin{align}
C^{2}=\frac{\sqrt{\frac{a_{0}M_{1}}{G}}}{4\pi\overline{\rho_{DM}(t)}}.
\end{align}
It is worth noting that certain galaxies, as observed in astronomy \cite{Comeron:2023dop, PinaMancera:2021wpc, vanDokkum:2018vup, vanDokkum:2022zdd, Danieli:2019zyi}, lack dark matter. An explanation for this phenomenon could be that their characteristic acceleration $a_{0}$ and mass $M_{1}$ are small enough to result in $C^{2}\sim0$. In such galaxies, the inflaton distribution appears to be relatively uniform and sparse so that the scale function of extra dimensions $\Phi(x^{\mu})\approx 1$. The relativistic contribution at the leading order to Newton's gravity law discussed in this subsection differs from the one presented in \cite{Das:2023zvp}. The results outlined in \eqref{modifiedpotential1} indicate an effective dark matter density profile
\begin{align}
\rho_{\text{eff}}(r)=\phi^{2}_{2}(r)\overline{\rho_{DM}(t)},
\end{align}
distinct from the Navarro-Frenk-White profile given in \eqref{density0}. It's important to mention that our focus is solely on the tree-level contribution to Newton's gravity law. For a more precise determination of the effective dark matter density profile, higher-order quantum corrections for the four-point matter scattering process can be incorporated. These loop corrections will provide a more refined contribution to the Tully-Fisher relations.

\subsection{Dark Energy}
In the late universe, assuming $a(t)\approx 1$, we consider the constraints imposed on $\phi_{1}(t)$ and $\phi_{2}(r)$ by equations
\begin{align}
\ddot{\phi_{1}}(t)+3H\dot{\phi_{1}}(t)&=0,\label{constraint3}\\
-\frac{1}{r^{2}}\partial_{r}\left(r^{2}\frac{\partial\phi_{2}(r)}{\partial r}\right)+m_{\phi}^{2}\phi_{2}(r)&=0.
\end{align}
The solutions to these equations representing dark energy are given by
\begin{align}
\phi_{1}(t)&=C_{1}+C_{2}a^{-3}(t),\\
\phi_{2}(r)&=C_{3}\frac{e^{-m_{\phi}r}}{r},
\end{align}
where $C_{1},C_{2},C_{3}$ are constants. The the function $F(t,\mathbf{x})$ are described by
\begin{align}
\left\langle F(t,\mathbf{x})\right\rangle_{t}=-\phi^{2}_{2}(r)\overline{\rho_{DE}(t)},
\end{align}
where $\phi_{0}=C_{1}+C_{2}$ as $a\simeq1$ and
 $\overline{\rho_{DE}(t)}=\frac{1}{2}m_{\phi}^{2}\phi_{0}^{2}=3m_{p}^{2}H^{2}\times\Omega_{DE}$. Here $\Omega_{DE}$ is the fraction parameter of dark energy.  After applying the Fourier transformation in \eqref{fourier00}, we discover that the spherically symmetric distribution of the inflaton generates a repulsive force between two massive objects. The Newton's gravity law for $M_{2}$ is given by
\begin{align}
F_{2}(R)=-\frac{GM_{1}M_{2}}{R^{2}}+\frac{4\pi GM_{2}\overline{\rho_{DE}(t)}C_{3}^{2}\left(1-e^{-2m_{\phi}R}\right)}{m_{\phi}R^{2}}.\label{gravity3}
\end{align}
This solution suggests that the extra dimensions can be interpreted as a spherically symmetric distribution of dark energy. This result aligns with the experimental constraints on the Yukawa gravitational potential \cite{Tan:2020vpf,Lee:2020zjt}, where the characteristic length $\lambda=\frac{\hbar}{2m_{\phi}c}\approx 4.04\times10^{-30}[\text{m}]$, but it is far beyond the current experimental accuracy to detect the gravitational effect \eqref{gravity3} at centimeter scale. It's worth noting that in this scenario, $w=\frac{p_{\text{DE}}}{\rho_{\text{DE}}}\approx-1$ indicating that the inflaton might be just one component of dark energy. However, the effective action \eqref{EHaction1} represents the tree-level effective action for eleven-dimensional supergravity. Higher-order derivatives of the inflaton, such as $\nabla^{\mu}\nabla^{\nu}{\phi}\nabla_{\mu}\nabla_{\nu}{\phi}$, can be found in the 1-loop order action of eleven-dimensional supergravity. We leave further discussions on the role of the inflaton as dark energy for future study.

\section{Conclusions and Discussion}
In this paper, we have discussed multiple potential methods for compactifying 11D supergravity, which can be represented as $\mathcal{M}^{3,1}\times T_{q_{1}}^{2}\times T_{q_{2}}^{2}\times T_{q_{3}}^{2}\times S^{1}$. As a low-energy effective theory derived from the ultraviolet-complete M-theory, the inherent completeness of quantum gravity in the ultraviolet regime does not forbid the existence of dark matter and dark energy in an FRW universe with four extended spacetime dimensions and seven extremely small compactified spatial dimensions. The dynamics of the inflaton, Standard Model of Particle Physics, dark matter, and dark energy can be all included in the effective action for the four-dimensional extended spacetime.  Meanwhile, the inflaton can be interpreted as the conformal vibrations of extra dimensions.

In our examples of driven potential \eqref{potential0}, the dynamics of sub-Planck scale extra dimensions govern the cosmological inflation that occurs in four-dimensional extended spacetime. When the scale function has minimum value $\Phi_{min}=-\frac{1}{n}$ and $s_{1},s_{2},s_{3}$ take values $s_{1}=\frac{1}{2},s_{2}=0,s_{3}\geq 1$, the theoretical predictions align with the observational range derived from the 2021 BICEP/Keck results, see Figure \ref{fig1}. The special case addressed in Subsection \ref{subectionA} satisfies all the observational constraints from both 2021 BICEP/Keck and 2018 Planck results. In this scenario, the geometric configuration of the extra dimensions is $S^{2}\times T_{2}^{2}\times T^{2}_{2}\times S^{1}$. As inflation progresses, the three tori undergo a contraction in size. Once the inflaton $\phi$ initiates its decay into gravitino $\psi_{A}$ and 3-form $A_{MNP}$, inflation exits. During this decay process, particles encompassed by the Standard Model of Particle Physics, as well as dark matter and dark energy, are generated. Then the extra dimensions become inactive and stabilize into a fixed size, where the standard sizes of the extra dimensions are around $10^{5}l_{p}$.

As mentioned in Section \ref{quantumEFF}, the quantum effects from extra dimensions are detectable in principle through high-energy experiments and astronomy observations. However, detecting the inflaton in accelerators is extremely challenging in practice. Nevertheless, there are other possible ways to indirectly infer the existence of extra dimensions, such as observations of the power spectrum of Cosmic Microwave Background (CMB) and the rotation curve of galaxies. On the one hand, in our compactification methods, the supersymmetry of gravity in four-dimensional extended spacetime is spontaneous breaking. This theoretical prediction aligns with experimental observations, as there is currently no conclusive evidence for superpartners of the graviton and the particles in the Standard Model of Particle Physics. On the other hand, the geometric configuration of the extra dimensions is reflected by the inflaton field, which can be interpreted as a component of dark matter or dark energy in the late universe. The modified Newton's gravity law, taking into account a spherically symmetric distribution of the inflaton, is presented in \eqref{gravity1}\eqref{gravity2}\eqref{gravity3}. The equation \eqref{gravity2}, which depicts a point particle surrounded by dark matter, can elucidate the Tully-Fisher relation given in \eqref{v4}. The range of the parameter $m_{2}$ in \eqref{gravity2} can be constrained through observations for galaxies. It is worth mentioning that our method to derive the properties of the dark sector is fixed on the Minkowski background and the backreaction sourced by matter to spacetime is neglected since the gravitational field is very weak. To study the case in the strong gravitational field, more precise calculations regarding the backreaction should be performed, which is left for future work.

\appendix
\section{Gauge Conditions}\label{Appendix1}
Because the gauge parameter $\theta_{MN}$ has 55 degrees of freedom, we can impose gauge conditions as follows:

1. For a given function symmetric tensor $\mathcal{T}_{\mu\nu}^{\text{(matter)}}[x^{\mu}]$, the equations
\begin{align}
T_{\mu\nu}^{(A,\psi)}[x^{A}]=\frac{m_{p}^{2}}{m_{p(11)}^{9}}\mathcal{T_{\mu\nu}^{\text{(matter)}}}[x^{\mu}],\label{gauge}
\end{align}
expect for $\mu=\nu=0$ case, restrict 9 degrees of freedom.

2. Particular block diagonal configuration of 11D bulk metric \eqref{11dmetric} restricts 46 degrees of freedom
\begin{align}
    G_{a_{i}A}&=0,(A\neq a_{i}\text{ and }i=1,2,3)\label{gaugeG1}\\
    G_{\mu A}&=0,(A\neq\mu)\label{gaugeG2}\\
    G_{10,A}&=0,(A\neq10)\label{gaugeG3}
\end{align}
By considering the gauge conditions \eqref{gauge}\eqref{gaugeG1}\eqref{gaugeG2}\eqref{gaugeG3}, boundary conditions, and the equations of motion for the graviton $G_{AB}$, gravitino $\psi_{A}$ and 3-from $A_{MNP}$ collectively, we can determine the solutions for the graviton, gravitino, and 3-form. These solutions are expressed as functions of the scale factor $a(t)$ and the scale function $\Phi(x^{\mu})$ by $\psi_{A}(a(t),\Phi(x^{\mu}),x^{A})$ and $A_{MNP}(a(t),\Phi(x^{\mu}),x^{A})$. To ensure the representation of spinors in four dimensions, the matrix $\Gamma^{M}$ can be expressed as the tensor product among a $4\times 4$ matrix $\hat{\gamma}^{\mu}$, three $2\times 2$ matrices $\hat{\gamma}^{a_{i}},i=1,2,3$, and one $2\times 2$ matrix $\hat{\gamma}^{10}$
\begin{align}
\Gamma^{\mu}&=\hat{\gamma}^{\mu}\otimes\mathds{1}_{2}\otimes\mathds{1}_{2}\otimes\mathds{1}_{2},\\
\Gamma^{a_{1}}&=\hat{\gamma}^{*}\otimes\hat{\gamma}^{a_{1}}\otimes\mathds{1}_{2}\otimes\mathds{1}_{2},\\
\Gamma^{a_{2}}&=\hat{\gamma}^{*}\otimes\hat{\gamma}_{1}^{*}\otimes\hat{\gamma}^{a_{2}}\otimes\mathds{1}_{2},\\
\Gamma^{a_{3}}&=\hat{\gamma}^{*}\otimes\hat{\gamma}_{1}^{*}\otimes\hat{\gamma}_{2}^{*}\otimes\hat{\gamma}^{a_{3}},\\
\Gamma^{10}&=\hat{\gamma}^{*}\otimes\hat{\gamma}_{1}^{*}\otimes\hat{\gamma}_{2}^{*}\otimes\hat{\gamma}^{10},
\end{align}
where $\hat{\gamma}^{*}=\hat{\gamma}^{0}\hat{\gamma}^{1}\hat{\gamma}^{2}\hat{\gamma}^{3},\hat{\gamma}_{1}^{*}=\hat{\gamma}^{5}\hat{\gamma}^{6}, \hat{\gamma}_{2}^{*}=\hat{\gamma}^{7}\hat{\gamma}^{8}.
$ Therefore, supersymmetry transformations introduce a 32-dimensional spinor gauge parameter denoted by $\epsilon_{\dot{\alpha}\dot{\alpha_{1}}\dot{\alpha_{2}}\dot{\alpha_{3}}}$, where $\dot{\alpha}=1,2,3,4$ and $\dot{\alpha}_{i}=1,2$ for $i=1,2,3$ are the four-dimensional spinor indices and two-dimensional spinor indices respectively. Given a specific classical solution of metric, as defined in \eqref{11dmetric}, we anticipate that its geometric configuration remains unaltered under the supersymmetry transformation outlined in \eqref{susytran1}\eqref{susytran2}\eqref{susytran3}. Consequently, this imposes certain gauge conditions, designated as
\begin{align}
    \delta g_{\mu\nu}=\delta g_{a_{1}b_{1}}=\delta h_{a_{2}b_{2}}=\delta \gamma_{a_{3}b_{3}}=0,\label{gaugeG}
\end{align}
where the operator $\delta(\cdots)$ is the variation with respect to supersymmetry. The supersymmetric transformation law for the metric $G_{AB}$ can be expressed as
\begin{align}
    \delta G_{AB}=\frac{1}{2}\eta_{ab}\bar{\epsilon}\left(\Gamma^{a}\psi_{A}e_{B}^{b}+e_{A}^{a}\Gamma^{b}\psi_{B}\right).\label{varG}
\end{align}
Additionally, its components are subject to gauge conditions \eqref{gaugeG}, denoted by
\begin{align}
\delta e_{\mu}^{a_{0}}&=\frac{1}{2}\bar{\epsilon}\Gamma^{a_{0}}\psi_{\mu}=0,a_{0}=1,2,3,4\label{varg}\\
\delta G_{a_{1}b_{1}}&=g_{a_{1}b_{1}}\delta\Phi,\label{varG1}\\
\delta G_{a_{2}b_{2}}&=0,\label{varG2}\\
\delta G_{a_{3}b_{3}}&=2g_{a_{3}b_{3}}\Phi\delta\Phi,\label{varG3}\\
\delta G_{10,10}&=-6\Phi^{-7}\delta\Phi.\label{varG10}
\end{align}
The gauge condition \eqref{varg} indicates the absence of gravitino, $\psi_{\mu}=0$, in the four-dimensional extended spacetime. Consequently, supersymmetry of gravity in the four-dimensional extended spacetime is spontaneous breaking. As there is no strong observable signal from supersymmetric field theory in 4D, we postulate that $\Phi$ has no superpartner, thus assuming $\delta\Phi=0$. Hence, the compactification methods spontaneously break all the $\mathcal{N}=32$ supersymmetries present in the eleven-dimensional bulk, where the classical solutions of \eqref{varg}\eqref{varG1}\eqref{varG2}\eqref{varG3}\eqref{varG10} are $\psi_{M}=0$. The non-trivial solution to the gauge parameter $\epsilon$ can be derived from the equation
\begin{equation}
\nabla_{M}\epsilon+\frac{\sqrt{2}}{288}\left(\Gamma_{\ \ \ \ \ \ \ M}^{ABCD}-8\Gamma^{BCD}\delta_{M}^{A}\right)F_{ABCD}\epsilon=0.\label{gaugeE}
\end{equation}
Note that there are trivial solutions $\epsilon=0$, when the 3-form takes trivial solution $A_{MNP}=0$. After imposing the gauge conditions \eqref{gauge}\eqref{gaugeG1}\eqref{gaugeG2} and \eqref{gaugeG3}\eqref{gaugeG}\eqref{gaugeE}, the only two remaining undetermined functions are the scale factor $a(t)$ and the scale function $\Phi(x^{\mu})$ in \eqref{11dmetric}.

In the four-dimensional extended spacetime, we anticipate that Einstein's gravity predominantly governs the laws of classical gravity as seen in \eqref{11dRicci}. Moreover, gauge conditions \eqref{gauge} imply that it is possible to obtain a four-dimensional effective action, in which the solutions to the equations of motion are represented by $\Phi(x^{\mu})$ and $a(t)$. While the equations of \eqref{gauge}, expect for $\mu=\nu=0$, have already been utilized as gauge conditions to determine the solution of 3-form, these equations do not establish the relation between $\Phi(x^{\mu})$ and $a(t)$.
To clarify the relationship between $\Phi(x^{\mu})$ and $a(t)$, we introduce an additional equation, which is the same as \eqref{gauge} but only for the case $\mu=\nu=0$. Subsequently, the solution for $\mathcal{T}^{\text{(matter)}}_{00}[\psi_{A},A_{MNP},a(t),\Phi(x^{\mu}),x^{\mu}]$ obtained from this equation uniquely defines the Lorentz-invariant symmetric tensor $\mathcal{T}^{\text{(matter)}}_{\mu\nu}[\psi_{A},A_{MNP},a(t),\Phi(x^{\mu}),x^{\mu}]$, thereby imposing 9 gauge conditions specified in \eqref{gauge}. These 9 gauge conditions, along with \eqref{gaugeG1}\eqref{gaugeG2}\eqref{gaugeG3}\eqref{gaugeG}\eqref{gaugeE}, constrain the solutions $\psi_{A}(a(t),\Phi(x^{\mu}),x^{A})$ and $A_{MNP}(a(t),\Phi(x^{\mu}),x^{A})$ in such a way that they determine $\mathcal{T}^{\text{(matter)}}_{\mu\nu}[a(t),\Phi(x^{\mu}),x^{\mu}]$ uniquely. This tensor can be derived by the equations of motion for $\Phi(x^{\mu})$ and $a(t)$ in four-dimensional extended spacetime. The equations \eqref{gauge} for $\mu,\nu=0,1,2,3$ serves to define an effective energy-stress tensor $\mathcal{T}_{\mu\nu}$ as
\begin{align}
\mathcal{T}_{\mu\nu}=\frac{23m_{p}^{2}}{2\Phi^{2}}\nabla_{\mu}\Phi\nabla_{\nu}\Phi+g_{\mu\nu}\left(-\frac{1}{2}\frac{23m_{p}^{2}}{2\Phi^{2}}\nabla_{\rho}\Phi\nabla^{\rho}\Phi-V(\Phi)\right)+\mathcal{T}_{\mu\nu}^{\text{(matter)}},
\end{align}
where $\mathcal{T}_{\mu\nu}^{\text{(matter)}}$ is the stress tensor of some four-dimensional matter. Hence, in the four-dimensional extended spacetime, the equation of motion \eqref{4dEinstein} implies 4D effective Einstein's equation
\begin{align}
    \mathcal{G}_{\mu\nu}=\frac{1}{m_{p}^{2}}\mathcal{T}_{\mu\nu},
\end{align}
and the associated effective action is given by
\begin{align}
S_{\text{eff}}=\int d^{4}x\sqrt{-g_{4}}\left[\frac{m_{p}^{2}}{2}R_{(4)}-\frac{1}{2}\frac{23m_{p}^{2}}{2\Phi^{2}}\nabla_{\mu}\Phi\nabla^{\mu}\Phi-V(\Phi)+\mathcal{L}_{\text{matter}}\right],\label{Effaction}
\end{align}
where
\begin{align}
\mathcal{T_{\mu\nu}^{\text{(matter)}}}=-2\frac{\delta\mathcal{L}_{\text{matter}}}{\delta g^{\mu\nu}}+g_{\mu\nu}\mathcal{L}_{\text{matter}}
\end{align}
The gravitino field $\psi_{A\dot{\alpha}\dot{\alpha_{1}}\dot{\alpha_{2}}\dot{\alpha_{3}}}$ gives rise to 56 fermion fields $\psi_{i\dot{\alpha}\dot{\alpha}_{1}\dot{\alpha_{2}}\dot{\alpha_{3}}}$ for $i=4,5,\cdots,10$ in the 4D extended spacetime, which contains the fermion fields in the Standard Model of Particle Physics. Additionally, there are 21 spin-1 boson fields $A_{\mu i j}$, 35 spin-0 scalar fields $A_{i j k}$, and 7 spin-0 tensor fields $A_{\mu\nu k}$ for $i,j,k=4,5,\cdots,10$. These fields contain ordinary matter, dark matter, and dark energy. Then, the lagrangian for matter in the 4D extended spacetime can encompass ordinary matter from the Standard Model of Particle Physics $\mathcal{L}_{\text{SM}}$, as well as dark matter $\mathcal{L}_{\text{DM}}$ and dark energy $\mathcal{L}_{\text{DE}}$,
\begin{align}
   \mathcal{L}_{\text{matter}}=\mathcal{L}_{\text{SM}}+\mathcal{L}_{\text{DM}}+\mathcal{L}_{\text{DE}},
\end{align}
and the associated effective action for the 4D extended spacetime is \eqref{Effaction}. This allows us to derive a cosmological inflation model based on a specified driven potential $V(\Phi)$, where the potential can be derived by integrating out the coordinates dependence of extra dimensions from the 11D supergravity action \eqref{sugra}.

\section{Properties of Potential}\label{Appendix}

We are going to prove the spectral index $n_{s}$ and tensor-to-scalar ratio $r$ are independent of the factor $(q_{3}-1)$ and parameter $n$ for the potential\eqref{potential0}. Perform a rescaling by
\begin{align}
\widetilde{n}&=e^{\alpha}n,\\
\widetilde{\Phi}&=e^{-\alpha}\Phi,\\
\widetilde{\phi}&=-\alpha/\sqrt{\lambda^{-1}}+\phi,
\end{align}
where the parameter $\alpha=-\frac{1}{2s_{3}}\ln k, k\in \mathbb{N_{+}}$. The potential differs only by a factor
\begin{align}
\widetilde{U}(\widetilde{\Phi})=U(\widetilde{\Phi})=e^{2s_{3}\alpha}U(\Phi).
\end{align}
The slow-roll parameter $\epsilon(\phi)$ transforms as
\begin{align}
\widetilde{\epsilon}(\widetilde{\phi})=\frac{m_{p}^{2}}{2}\left(\frac{\widetilde{U}_{\widetilde{\phi}}}{\widetilde{U}}\right)^{2}=\frac{m_{p}^{2}}{2}\left(\frac{e^{2s_{3}\alpha}U_{\phi}d\phi/d\widetilde{\phi}}{e^{2s_{3}\alpha}U}\right)^{2}=\epsilon(\phi).
\end{align}
Hence, we find the parameter $\alpha$ does not affect the slow-roll parameter $\epsilon(\phi)$.  This is the same for $\eta(\phi)$. Therefore the spectral index $n_{s}$ and tensor-to-scalar ratio $r$ are independent of $\alpha$. As a result, we can select $\widetilde{n}=e^{\alpha}$ such that $n=1$. The genus of the third torus can be chosen to be $q_{3}=1+k$, then the potential $\widetilde{U}(\widetilde{\Phi})$ dropped factor $(q_{3}-1)$.

\section{Exit Inflation}\label{Appendix3}
There exist additional matter terms within the four-dimensional effective action of 11D Supergravity \eqref{Effaction}. Because the 11D metric $G_{AB}$ couples to gravitino $\psi_{A}$ and 3-form
$A_{MNP}$, the inflaton $\phi$, associated
to the size function $\Phi$ of extra dimensions, will couple to some components of $\psi_{A}$
and $A_{MNP}$ after compactification. When the inflaton $\phi$ begins to decay into the gravitino and the 3-form, the inflation exits. In this period, we keep the supersymmetry gauge condition $\delta g_{\mu\nu}=0$ and release $\delta g_{a_{1}b_{1}},\delta h_{a_{2}b_{2}},\delta \gamma_{a_{3}b_{3}}\neq0$, see \ref{Appendix1} for more details about the supersymmetry gauge condition during inflation period. This decay process gives rise to a non-zero $\mathcal{L}_{\text{matter}}$ term in the four-dimensional effective action. For example, matter in the universe can be generated by introducing the interactions between the inflaton and the Higgs field $H$, expressed as $\xi\phi^{2} H^{2}$, into the Lagrangian $\mathcal{L}_{\text{matter}}$.

\section*{ACKNOWLEDGEMENTS}
We would like to thank Chi-Ming Chang, Zhengyuan Du, Wen-Xin Lai, Jian Xin Lu, Wei Song, Yi-xiao Tao, Dongjian Wu, and Tianqing Zhu for useful discussions. The work is partially supported by the national key research and development program of China NO. 2020YFA0713000.

\bibliographystyle{apsrev4-2}

\bibliography{sample.bib}

\end{document}